\PassOptionsToPackage{table,xcdraw}{xcolor}
\documentclass[sigconf,natbib=true]{acmart}
\AtBeginDocument{%
  }

\usepackage{booktabs}
\usepackage{multirow}
\usepackage{graphicx}
\usepackage{enumitem}
\usepackage{tabularx}

\newtheorem{lemma}{Lemma}

\newcommand{\ie}{\emph{i.e., }}
\newcommand{\eg}{\emph{e.g., }}

\newcommand{\cf}{\emph{cf. }}
\newcommand{\aka}{\emph{a.k.a. }}


\copyrightyear{2025}
\acmYear{2025}
\setcopyright{acmlicensed}\acmConference[SIGIR '25]{Proceedings of the 48th International ACM SIGIR Conference on Research and Development in Information Retrieval}{July 13--18, 2025}{Padua, Italy}
\acmBooktitle{Proceedings of the 48th International ACM SIGIR Conference on
Research and Development in Information Retrieval (SIGIR '25), July 13--18, 2025, Padua, Italy}
\acmDOI{10.1145/3726302.3730041}
\acmISBN{979-8-4007-1592-1/2025/07}

\begin{document}

\title{MSL: Not All Tokens Are What You Need for Tuning LLM as a Recommender}

\author{Bohao Wang}
\email{bohao.wang@zju.edu.cn}
\authornote{This work was done during an internship at OPPO Research Institute.}
\authornotemark[3]
\authornotemark[4]
\orcid{0009-0006-8264-3182}
\affiliation{%
  \institution{Zhejiang University}
  \city{Hangzhou}
  \country{China}
}

\author{Feng Liu}
\email{liufeng4hit@gmail.com}
\affiliation{%
  \institution{OPPO Research Institute}
  \city{Shenzhen}
  \country{China}
}

\author{Jiawei Chen}
\email{sleepyhunt@zju.edu.cn}
\orcid{0000-0002-4752-2629}
\authornote{Corresponding author.}
\authornote{State Key Laboratory of Blockchain and Data Security, Zhejiang University.}
\authornote{College of Computer Science and Technology, Zhejiang University.}
\authornote{Hangzhou High-Tech Zone (Binjiang) Institute of Blockchain and Data Security.}
\affiliation{%
  \institution{Zhejiang University}
  \city{Hangzhou}
  \country{China}
}

\author{Xingyu Lou}
\email{louxingyu@oppo.com}
\affiliation{%
  \institution{OPPO Research Institute}
  \city{Shenzhen}
  \country{China}
}

\author{Changwang Zhang}
\email{changwangzhang@foxmail.com}
\affiliation{%
  \institution{OPPO Research Institute}
  \city{Shenzhen}
  \country{China}
}

\author{Jun Wang}
\email{junwang.lu@gmail.com}
\affiliation{%
  \institution{OPPO Research Institute}
  \city{Shenzhen}
  \country{China}
}

\author{Yuegang Sun}
\email{bulutuo@i-i.ai}
\affiliation{%
  \institution{Intelligence Indeed}
  \city{Hangzhou}
  \country{China}
}

\author{Yan Feng}
\email{fengyan@zju.edu.cn}
\orcid{0000-0002-3605-5404}
\authornotemark[3]
\authornotemark[4]
\affiliation{%
  \institution{Zhejiang University}
  \city{Hangzhou}
  \country{China}
}

\author{Chun Chen}
\email{chenc@zju.edu.cn}
\orcid{0000-0002-6198-7481}
\authornotemark[3]
\authornotemark[4]
\affiliation{%
  \institution{Zhejiang University}
  \city{Hangzhou}
  \country{China}
}

\author{Can Wang}
\email{wcan@zju.edu.cn}
\orcid{0000-0002-5890-4307}
\authornotemark[3]
\authornotemark[5]
\affiliation{%
  \institution{Zhejiang University}
  \city{Hangzhou}
  \country{China}
}

\renewcommand{\shortauthors}{Bohao Wang et al.}

\begin{abstract}
Large language models (LLMs), known for their comprehension capabilities and extensive knowledge, have been increasingly applied to recommendation systems (RS). Given the fundamental gap between the mechanism of LLMs and the requirement of RS, researchers have focused on fine-tuning LLMs with recommendation-specific data to enhance their performance. Language Modeling Loss (LML),  originally designed for language generation tasks, is commonly adopted. However, we identify two critical limitations of LML: 1) it exhibits significant divergence from the recommendation objective; 2) it erroneously treats all fictitious item descriptions as negative samples, introducing misleading training signals.

To address these limitations, we propose a novel \textbf{Masked Softmax Loss (MSL)} tailored for fine-tuning LLMs on recommendation. MSL improves LML by identifying and masking invalid tokens that could lead to fictitious item descriptions during loss computation. This strategy can effectively avoid the interference from erroneous negative signals and ensure well alignment with the recommendation objective supported by theoretical guarantees. During implementation, we identify a potential challenge related to gradient vanishing of MSL. To overcome this, we further introduce the temperature coefficient and propose an \textbf{Adaptive Temperature Strategy (ATS)} that adaptively adjusts the temperature without requiring extensive hyperparameter tuning. Extensive experiments conducted on four public datasets further validate the effectiveness of MSL, achieving an average improvement of 42.24\% in NDCG@10. The code is available at \url{https://github.com/WANGBohaO-jpg/MSL}.

\end{abstract}


% \received{20 February 2007}
% \received[revised]{12 March 2009}
% \received[accepted]{5 June 2009}

%%
%% This command processes the author and affiliation and title
%% information and builds the first part of the formatted document.
\maketitle

\section{Introduction}
% Large Language Models (LLMs) have showcased exceptional capabilities in content comprehension, generation, and semantic reasoning, driving transformative advancements in artificial intelligence. Recently, their application in the domain of Recommendation Systems (RS) has garnered significant attention. A prominent approach involves leveraging LLMs as the backbone for recommendation tasks, utilizing their capacity to analyze users' historical interactions to infer preferences and predict future behaviors. Early efforts primarily focused on designing suitable prompts to guide LLMs in understanding and executing recommendation tasks. However, the inherent misalignment between the pre-training objectives of LLMs and the specific requirements of recommendation tasks often leads to suboptimal performance in such direct applications. To address this, subsequent research has explored instruction tuning, where LLMs are fine-tuned on recommendation-specific datasets, resulting in notable performance improvements. 

Large Language Models (LLMs) have showcased exceptional capabilities in content comprehension and leveraging extensive knowledge, thereby catalyzing a revolution in artificial intelligence \cite{achiam2023gpt}. Recently, LLMs have been extensively applied in the field of Recommender Systems (RS) \cite{wu2024survey}. A prominent strategy involves directly leveraging LLMs as recommenders --- organizing users' historical interactions as language prompts and instructing LLMs to deduce users' preferences for predicting future interactions \cite{hou2024large,wang2023zero,gao2023chat,wang2024recommend,liu2024once}. This paradigm has demonstrated enhanced few-shot ability \cite{hou2024large, wang2023zero}, generalization \cite{kolb2024enhancing}, explainability \cite{gao2023chat}, and impressive recommendation performance \cite{bao2023tallrec}.

\begin{figure}[t]
  \centering
  \includegraphics[width=\linewidth]{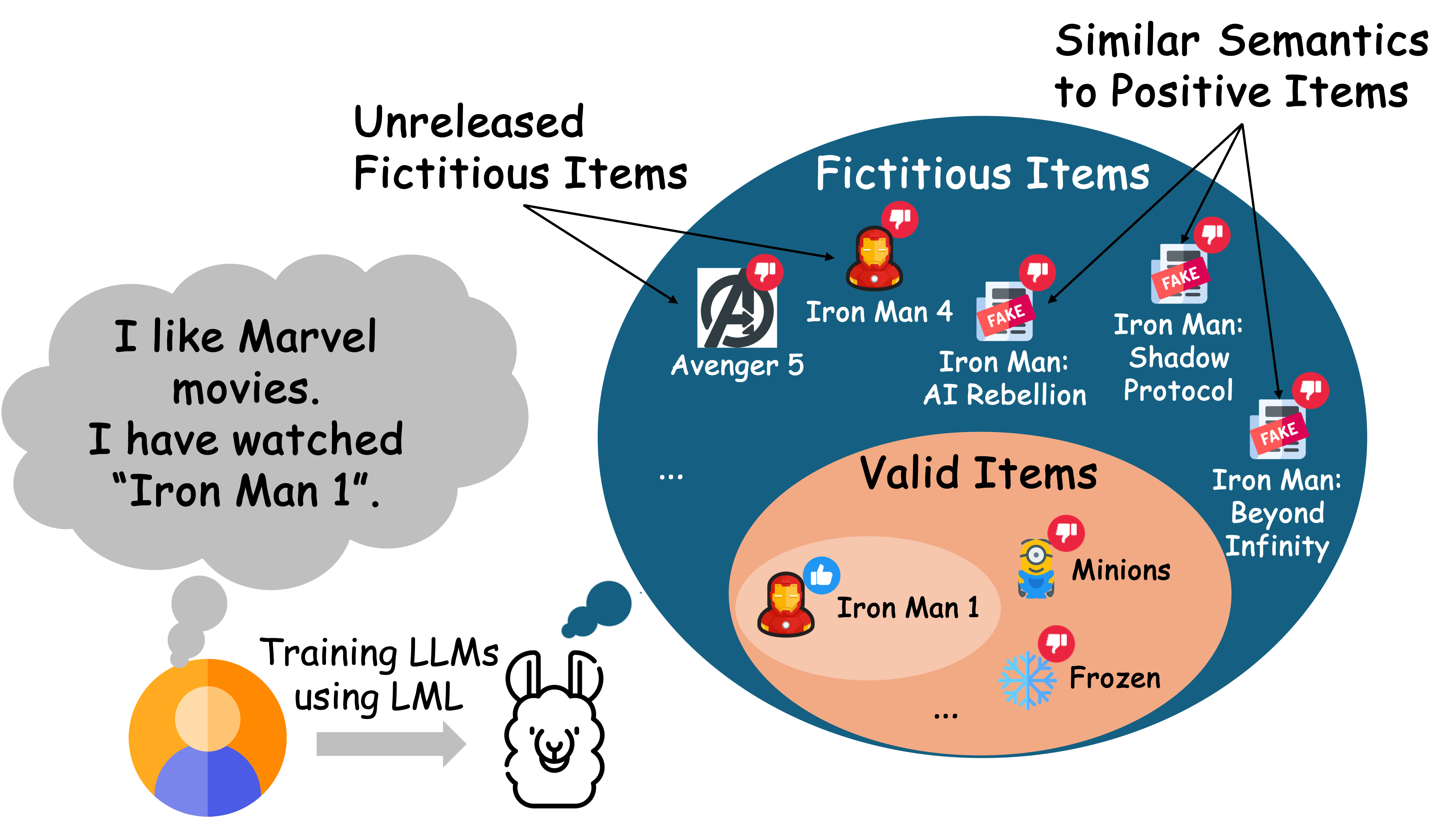}
  % \caption{Language modeling loss will mistakenly treat all fictional items as negative items. However, some of these items may share semantic similarities with the items the user already favors (\eg "Iron Man: AI Rebellion"). Besides, certain items may be considered fictional simply because they have not yet been released (\eg "Iron Man 4"). These cases represent incorrect negative sampling signals.}
  \caption{Language modeling loss can erroneously treat all fictional items as negative items. However, some of these items may exhibit semantic similarities to positive items (\eg "Iron Man: AI Rebellion"). Additionally, certain items might be fictional simply because they have not yet been released (\eg "Iron Man 4"). These cases represent incorrect negative signals.}
  \Description{}
  \label{fake_items}
\end{figure}

To fully unlock the potential of LLMs in recommendation, supervised fine-tuning (SFT) is commonly applied for LLM-based recommenders \cite{bao2023tallrec, bao2023bi, liao2024llara, kim2024large, gao2024end, zheng2024adapting}. These methods typically structure users' historical interactions as prompts, paired with descriptions of positive items as target responses, and fine-tune LLMs using a Language Modeling Loss (LML) \cite{radford2018improving}. This loss, inherited from language generation tasks and expressed as a token-wise softmax loss, augments the probability (\ie logits) of tokens representing positive items while penalizing the logits of other generated content. However, we argue that this objective has significant limitations in the recommendation scenario:
\begin{itemize}[left=5pt]
    \item \textbf{Significant Divergence from the Recommendation Objective:} RS aims for personalized ranking performance (\eg higher NDCG), prioritizing positive items over negative ones. LML deviates significantly from this ranking objective. Through extensive theoretical and empirical analyses, we find that optimizing LML primarily focuses on generating valid item descriptions that exist in the system, while providing limited guidance to help LLMs differentiate positive items from negative ones. This deviation significantly hinders the effectiveness of LML in recommendations.
    \item  \textbf{Improper Negative Signals:} Language modeling loss implicitly considers all other generated item descriptions as negative, including valid negative items that the user has not interacted with and fictitious items that do not exist in the RS. This treatment is flawed as it is improper to hypothesize that the user dislikes these fictitious items. In fact, some fictitious items may share similar semantic with the positive items, whose contents may be favored by users. 
    As shown in Figure \ref{fake_items}, a typical fan of the Marvel Universe who has watched "Iron Man 1" would likely enjoy a fictitious movie such as "Iron Man: AI Rebellion". Blindly treating all such fictitious items as negative could confuse the LLM in capturing user preferences.
\end{itemize}

\begin{figure}[t]
  \centering
  \includegraphics[width=\linewidth]{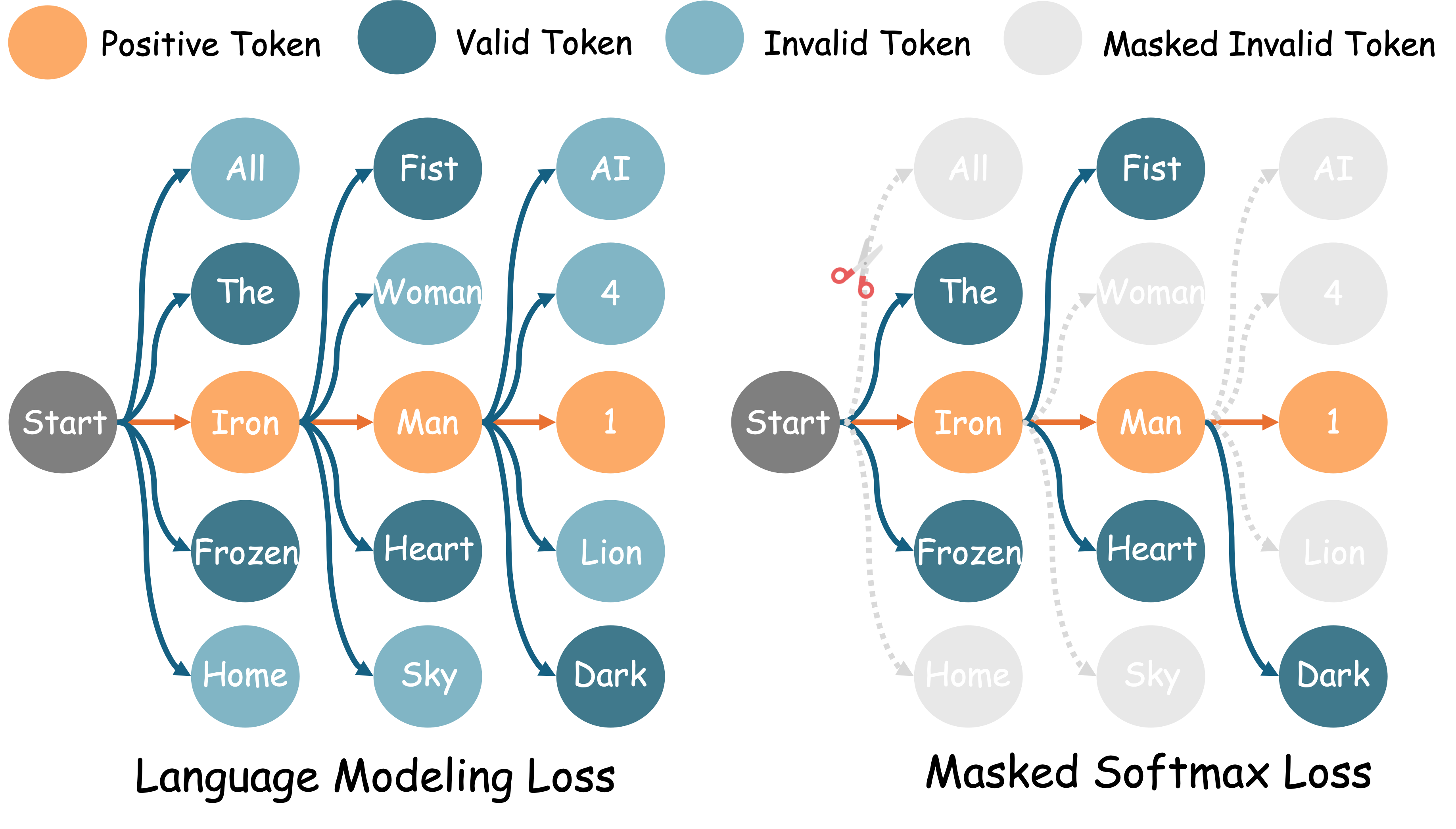}
  \caption{The schematic diagram illustrates Language Modeling Loss (LML) and Masked Softmax Loss (MSL).}
  \Description{}
  \label{method}
\end{figure}

To tackle these limitations, we introduce a novel loss function, termed \textbf{Masked Softmax Loss (MSL)}, specifically designed for fine-tuning LLMs as recommenders. MSL employs a masking mechanism that prevents penalization of fictitious item descriptions. As illustrated in Figure \ref{method}, this mechanism can be implemented efficiently at the token level by masking the invalid tokens in the softmax calculation that correspond fictitious items. Our theoretical analyses further demonstrate the close connection of MSL with the ranking objective, serving as a tight upper bound of the NDCG metrics. 
% During the inference period, this masking strategy can also be employed to constrain the generated content to valid items only, effectively addressing the issue of hallucination where LLMs generate fictitious items.

% To address the limitations of SL, we introduce a novel loss function, termed \textbf{Masked Softmax Loss (MSL)}, specifically designed for training LLMs as recommenders. MSL first constructs a Trie tree to identify the valid tokens at each position within the trainable token sequences, then restricting the negative samples in SL to valid tokens instead of using all tokens as negative samples. Based on this simple modification, MSL effectively establishes consistency with the recommendation objective. We provide theoretical proof that MSL offers a tighter lower bound on NDCG compared to SL. Moreover, the construction of the Trie tree is highly efficient and only needs to be performed once during the data preprocessing stage. As a result, MSL incurs no additional computational overhead during model training compared to SL.

Despite its theoretical advantages, MSL may encounter gradient vanishing issues during practical application. This arises from the reduced number of tokens in the softmax denominator, which can lead to particularly small gradients and loss values. A simple and effective strategy to address this is the introduction of an additional hyperparameter, temperature, in the softmax function to modulate its values. While effective, this approach requires tedious and time-consuming hyperparameter tuning, which is unsatisfactory for LLM-based recommendations. To address this challenge, we propose an Adaptive Temperature Strategy (ATS). By examining the gradient of MSL and the role of temperature, we derive an adaptive configuration based on the average number of valid tokens in the dataset. This strategy effectively mitigates gradient vanishing in MSL without requiring extensive hyperparameter tuning.

% Despite its theoretical advantages, MSL may encounter gradient vanishing issues during practical training. To address this, we propose an \textbf{Adaptive Temperature Strategy (ATS)}, which dynamically adjusts the temperature parameter during training to stabilize the probability values of positive samples, thereby preventing gradient vanishing caused by excessively high probabilities. This approach not only mitigates gradient vanishing but also reduces the need for extensive parameter tuning, a common challenge in training LLMs. Together, MSL and ATS offer a robust and efficient framework for improving recommendation performance in LLMs.

Lastly, in terms of aligning LLM with the ranking objective, the most relevant work is the recently proposed S-DPO \cite{chen2024softmax}, which integrates Direct Preference Optimization (DPO) \cite{rafailov2024direct} in LLM-based recommendation. However, S-DPO exhibits several limitations: 1) Suboptimal Performance: S-DPO still relies on LML to fine-tune the model and considers the fine-tuned LLM as a reference model for optimization. Given the inherent limitations of LML, the effectiveness of S-DPO is compromised. 2) Unstable Results: S-DPO requires negative item sampling, which can lead to training instability, especially in fine-tuning tasks with limited epochs. 3) High Computational Cost: S-DPO requires more training instances and epochs, resulting in significantly longer training times (approximately 4 times) compared to MSL.

% \begin{figure}[t]
%   \centering
%   \includegraphics[width=\linewidth]{method1.png}
%   \caption{The schematic diagram illustrates Language Modeling Loss (LML) and Masked Softmax Loss (MSL). The upward and downward arrows respectively indicate whether the optimization is to increase or decrease the logits of the token.}
%   \Description{}
%   \label{method}
% \end{figure}

% Currently, there is a noticeable lack of research focused on developing specialized loss functions tailored for recommendation systems. Among the existing methods, S-DPO represents a notable attempt to redesign loss functions for recommendation tasks. Building on the foundation of DPO, S-DPO introduces additional negative samples to improve the model's ability to capture ranking relationships among items. However, S-DPO suffers from two major limitations: (1) it is prone to converging to local optima when the number of negative samples is insufficient, and (2) its computational cost increases significantly as the number of negative samples grows. In fact, under certain conditions, it can be theoretically demonstrated that our proposed MSL is equivalent to the optimal form of S-DPO, where all items in the item set are treated as negative samples.

In summary, this work makes the following contributions:
\begin{itemize}[left=5pt]
\item We propose a novel loss function, Masked Softmax Loss (MSL), specifically tailored for fine-tuning large language models to effectively align with recommendation objectives.
\item We address the potential gradient vanishing issue of MSL by developing an adaptive temperature strategy that mitigates this issue without requiring hyperparameter tuning.
\item Extensive experiments on four real-world datasets demonstrate that the proposed MSL outperforms LML by a large margin (42.24\% on average in NDCG@10).
\end{itemize}

\begin{table}[t]
\centering
\caption{Prompt templates for implementing recommendation tasks (using Toy dataset as an example)}
\begin{tabularx}{\columnwidth}{@{}lX@{}}
    \toprule
    \multicolumn{2}{c}{\textbf{Instruction Input}} \\
    \midrule
    \textbf{Instruction:} & Given a list of toys the user has played before, please recommend a new toy that the user likes to the user. \\
    \cmidrule(l){1-2}
    \textbf{Input:} & The user has played the following toys before: "LeapFrog Discovery Ball", "Plush Elmo Knows Your Name", "Blokus Game", ... \\
    \midrule
    \multicolumn{2}{c}{\textbf{Instruction Output}} \\
    \midrule
    \textbf{Output:} & "MindWare Q-Ba-Maze Cool Colors" \\
    \bottomrule
\end{tabularx}
\label{tab:LLM4Rec_example}
\end{table}

\section{LLM-based Recommendation}
Referring to recent work \cite{liao2024llara, bao2023bi, na2024enhancing, bao2023tallrec, lin2024rella, zheng2024adapting}, this work also focuses on sequential recommendation, which holds notable practical significance by considering the temporal order of user behavior. Given a sequential recommender system with a user set $\mathcal{U}$ and an item set $\mathcal{V}$, let user's historical interactions be denoted as $S=\{s_1,s_2, ...\}$, where $s_i \in \mathcal{V}$ denotes the $i$-th interacted item in the sequence. The objective of sequential recommendation is to infer user preferences from $S$ and retrieve the positive item $p$ that the user will interact with next. This task is often conceptualized as a ranking problem, aiming to position the positive item $p$ higher in the ranking list. Consequently, ranking metrics such as NDCG are frequently adopted to evaluate recommendation performance.

Given the remarkable success of large language models (LLMs) across various domains \cite{wu2023visual, ouyang2022training, tang2024graphgpt}, integrating LLMs into recommendation systems has been extensively explored \cite{wu2024survey}. A prominent strategy is to directly leverage powerful LLMs as recommenders. As shown in Table \ref{tab:LLM4Rec_example}, this paradigm organizes users' historical interactions as language prompts $x$, typically consisting of the descriptions (\eg titles) of the items in $S$ and the description of the recommendation task. This prompt is then used to instruct the LLMs to predict the item (descriptions) that the user is most likely to interact with. 

Since LLMs are typically not pre-trained on recommendation data, supervised fine-tuning is necessary to align LLMs with the recommendation task. This strategy pairs the prompts $x$ and the description of the target positive item $y^p$ as a training instance $(x, y^p)$, and optimizes LLMs with the following \textbf{Language Modeling Loss (LML)}:

\begin{equation}
\begin{aligned}
 \mathcal{L}_{LML}(x, y^p; \theta) &= - \log P_{\theta} \left(y^p \mid x\right)
= \sum_{t=1}^{|y^p|} - \log P_{\theta} \left(y^p_t \mid x, y^p_{<t}\right) \\
&= \sum_{t=1}^{|y^p|} - \log \frac{\exp(f_\theta(y_t^p|x,y^p_{<t}))}{\sum_{z \in \mathcal{Z}} \exp(f_\theta(z|x,y^p_{<t}))}
\end{aligned}
\end{equation}
where $y^p_t$ denotes the $t$-th token of the positive item description $y^p$, and $y^p_{<t}$ represents the token sequence preceding $y^p_t$. The set $\mathcal{Z}$ corresponds to the entire vocabulary of tokens in the LLM, and $f_\theta(y^p_t \mid x, y^p_{<t})$ denotes the logit of the token $y^p_t$ predicted by LLMs, where $\theta$ denotes the parameters of LLMs. For simplicity, we use \(f_\theta(y^p_t)\) \textit{(or \(f_\theta(z)\))} to represent \(f_\theta(y^p_t \mid x, y^p_{<t})\) \textit{(or \(f_\theta(z \mid x, y^p_{<t})\))}, 
% and \(P_{\theta}(y^p)\) \textit{(or \(P_{\theta}(y^p_t)\))} to denote \(P_{\theta}(y^p \mid x)\) \textit{(or \(P_{\theta}(y^p_t \mid x, y^p_{<t})\))}.
and \(P_{\theta}(y^p_t)\) to denote \(P_{\theta}(y^p_t \mid x, y^p_{<t})\).

The language modeling loss is directly inherited from language generation tasks, aiming to maximize the probability of the descriptions of positive items over the whole generative content space. It can be expressed in a token-wise manner with softmax loss, which augments the logits of the tokens representing positive items (numerator), while decreasing the logits of the other tokens in the vocabulary (denominator).

The notation table is presented in Table \ref{notation}.

\begin{table}[t]
    \centering
    \caption{Notations in the paper.}
    \begin{tabular}{ll}
    \toprule 
    \textbf{Notations}&\textbf{Descriptions}\\
    \midrule 
    $\mathcal{U}$ & user set \\
    $\mathcal{V}$ &  item set \\
    
    $S$ &  the user historical interaction sequence \\
    $p$ &  the positive item of the sequence $S$ \\
    $x$ &  the input prompt of the sequence $S$ \\
    $y^v$ &  the description of the item $v$ \\
    $y^v_t$ &  $t$-th token of $y^v$ \\

    $\mathcal{Z}$ &  the vocabulary of LLM \\
    $\mathcal{Z}_{valid}(y^v_{<t})$ & valid tokens for a given prefix $y^v_{<t}$ \\ 
    % $\mathcal{Z}'$ &  negative tokens in MSL \\
    
    $\theta$ &  the model parameter \\
    $f_\theta$ &  logits output by the model \\
    $P_\theta(y^v_t)$ &  the probability of \( y^v_t \) over $\mathcal{Z}$ \\
    $P_\theta^{valid}(y^v_t)$ &  the probability of \( y^v_t \) over \( \mathcal{Z}_{valid}(y^v_{<t}) \) \\
    % $\eta$ & the hyperparameter in ATS \\
    % $\tau$ & the temperature in MSL \\
    % $m$ & the average number of valid tokens \\
    % $\mu$ & the mean of logits \\
    % $\sigma^2$ & the variance of logits \\
    $\mathcal{L}_{LML}$ &  language modeling loss \\
    $\mathcal{L}_{MSL}$ &  masked softmax loss \\
    % $\hat{l}$ &  the average token length of the item description \\
    \bottomrule
    \label{notation}
    \end{tabular}
\end{table}

\section{Analyses on Language Modeling Loss}
\label{sec:analyse}
While language modeling loss is commonly used for fine-tuning LLMs as recommenders, we argue that it still suffers from the following limitations:

\textbf{Limitation 1: Significant Divergence from the Recommendation Objective.} Recommender systems aim to retrieve positive items from the valid item set in the system. In contrast, LML aims to retrieve positive item descriptions from the entire language space that LLMs could generate. It's important to note that the language space contains descriptions of both valid items in the system and fictitious items imagined by LLMs. This causes the objective of LML to deviate significantly from the recommendation objective.

To better understand this deviation, we can decompose LML into two components:
\begin{align}
\label{eq:decompose}
\mathcal{L}_{LML}(x, y^p;\theta) & = \mathcal{L}_{LML}^1(x, y^p;\theta) + \mathcal{L}_{LML}^2(x, y^p;\theta) \\
\mathcal{L}_{LML}^1(x, y^p;\theta) & = \sum_{t=1}^{|y^p|} \underbrace{- \log \frac{\sum_{z \in \mathcal{Z}_{valid}(y^p_{<t})} \exp(f_\theta(z))}{\sum_{z \in \mathcal{Z}} \exp(f_\theta(z))}}_{\text{Lifting valid items over invalid items}} \\
\mathcal{L}_{LML}^2(x, y^p;\theta) & = \sum_{t=1}^{|y^p|} \underbrace{- \log \frac{\exp(f_\theta(y_t^p))}{\sum_{z \in \mathcal{Z}_{valid}(y^p_{<t})} \exp(f_\theta(z))}}_{\text{Lifting the positive items over negative items}}
\end{align}
where $z \in \mathcal{Z}_{valid}(y^p_{<t})$ denotes a valid token, ensuring that the combined language contents $[y^p_{<t}, z]$ can be a prefix of any valid item description. A similar definition applies to invalid tokens, $z \notin \mathcal{Z}_{valid}(y^p_{<t})$, which would make the generated contents fall outside the scope of valid items' descriptions. For simplicity, we use \(\mathcal{Z}_{valid}\) to denote \(\mathcal{Z}_{valid}(y^p_{<t})\) in the following text.

\begin{figure}[t]
  \centering
  \includegraphics[width=\linewidth]{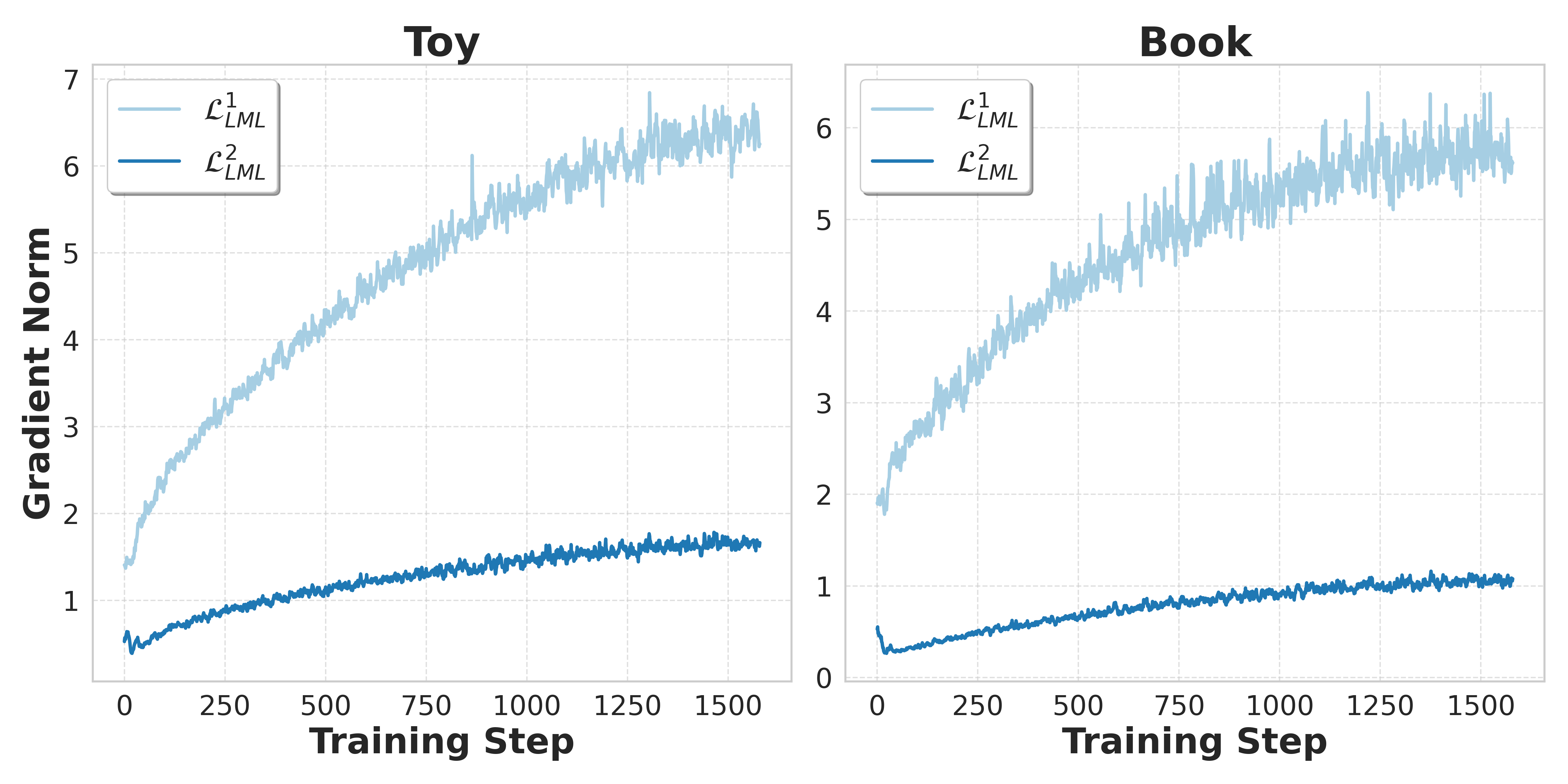}
  \caption{The gradient norms of the two components of language modeling loss during the model training.}
  \Description{}
  \label{grad_norm}
\end{figure}

\begin{figure}[t]
  \centering
  \includegraphics[width=\linewidth]{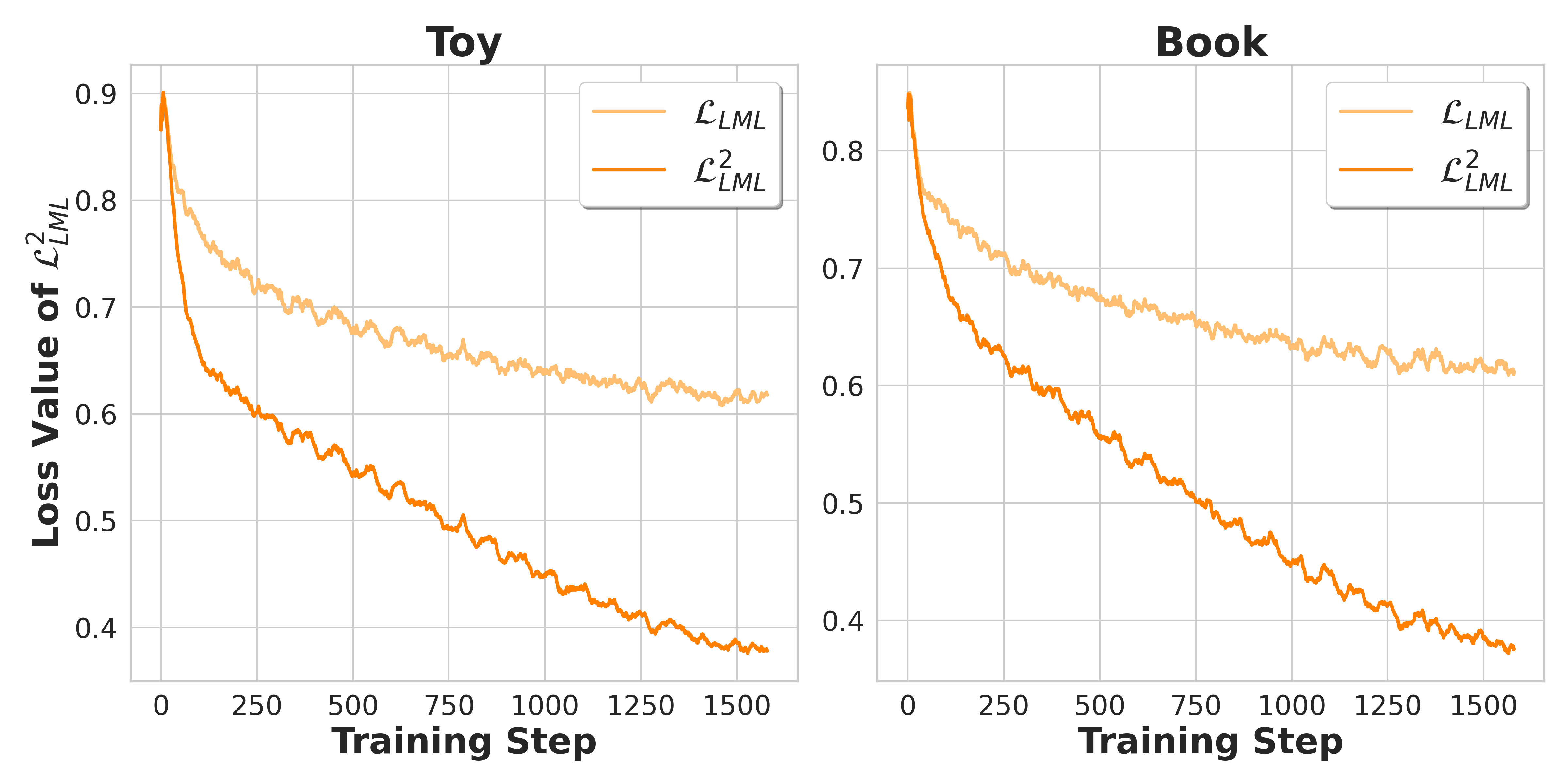}
  \caption{The variation of \(\mathcal{L}_{LML}^2\) during the training when optimizing with \(\mathcal{L}_{LML}\) vs. \(\mathcal{L}_{LML}^2\).}
  \Description{}
  \label{loss}
  % \vspace{-0.6cm}
\end{figure}

Language modeling loss has two-fold effects: 1) $\mathcal{L}_{LML}^1$ increases the logits of valid tokens while penalizing the invalid tokens. This component would lift the probability of valid items over fictitious items, guiding the LLMs towards outputting a valid item description. 2) $\mathcal{L}_{LML}^2$ increases the logits of the positive token (\ie $y^p_t$) and penalizes the logits of negative tokens (\ie $\mathcal{Z}_{valid} \setminus y^p_t$). This component would lift the probability of positive items over negative items, which aligns with the recommendation objective. We will also prove the close theoretical connection of this component with the NDCG metrics in the next section (lemma \ref{lemma_ndcg}). 

The above decomposition illustrates the differences and connections between LML and the recommendation objective. While the recommendation objective serves as one component of LML, we empirically find that optimizing LML is ineffective, as the gradient is dominated by $\mathcal{L}_{LML}^1$. Figure \ref{grad_norm} illustrates this point, showing the norm of the gradient from two components on typical datasets Toy and Book. It can be observed that $\mathcal{L}_{LML}^1$ exerts an overwhelming effect on the training, hindering the convergence of $\mathcal{L}_{LML}^2$. 
To further demonstrate this point, we conduct another experiment as shown in Figure \ref{loss}, where we visualize the training loss of $\mathcal{L}^2_{LML}$ when we optimize $\mathcal{L}_{LML}$ or $\mathcal{L}_{LML}^2$ only for comparison. As can be seen, the training loss when we directly optimize $\mathcal{L}_{LML}^2$ decreases quickly, while the loss drop under optimizing $\mathcal{L}_{LML}$ seems hindered. These analyses demonstrate the ineffectiveness of leveraging LML in improving recommendation performance.

\textbf{Limitation 2: Improper Negative Signals.} From Eq.(\ref{eq:decompose}), we find that LML penalizes the logits of invalid tokens, implicitly considering all fictitious items that are not exist on the system as negative items. However, this treatment is flawed as it is improper to hypothesize that the user dislikes these fictitious items. In fact, some fictitious items may share semantic similarities with positive items and could potentially align with user preferences.
% For example, if a user has watched the movie "xx", it is improper to hypothesize that the user dislikes a highly similar movie named "xx" even though it is imagined. Similarly, considering a user is a fan of the Marvel Universe and has watched movies like "Iron Man" and "Spider-Man", it is highly likely that the user would like the movie "xx" if it existed in the system. 
To illustrate this issue, consider the example of a typical fan of the Marvel Universe who enjoys the movie "Iron Man 1" as shown in Figure \ref{fake_items}. Some of the fictitious items may share semantic similarities with the positive items and may be favored by users (\eg "Iron Man: AI Rebellion"). Additionally, certain items are fictional simply because they have not yet been released (\eg "Iron Man 4").
As such, blindly treating all such fictitious items as negative could confuse the LLM, giving incorrect signals for capturing user preference.

% Please add the following required packages to your document preamble:
% \usepackage{booktabs}
\begin{table}[]
\centering
\caption{The deviation of NDCG@10 under multiple random seeds.}
\label{tab:deviation}
\begin{tabular}{@{}lcc@{}}
\toprule
Method & Toy                 & Book                 \\ \midrule
S-DPO  & $0.0219 \pm 0.0036 (16.4\%)$ & $0.0124 \pm 0.0019 (15.3\%)$  \\
MSL    & $0.0294 \pm 0.0006 (2.0\%)$ & $ 0.0175 \pm 0.0005 (2.9\%)$ \\ \bottomrule
\end{tabular}
\end{table}

\textbf{Analyses on S-DPO}.
While S-DPO \cite{chen2024softmax} leverages direct preference optimization to enhance LLM-based recommendation, it still suffers from the following limitations:

\begin{itemize}[left=5pt]
\item \textbf{Suboptimal Performance.} S-DPO can not address the aforementioned limitations inherent in LML. S-DPO still relies LML to fine-tune LLMs, which would be utilized as a reference model for further DPO optimization. Given the inherent limitations of LML, the effectiveness of S-DPO
is compromised (\cf Section \ref{sec:cp_exp}).
% LML to fine-tune the model and
% considers the fine-tuned LLM as a reference model for optimization.

% S-DPO constrains the output distribution of the fine-tuned model to not deviate significantly from that of the reference model (SFT by LML). This mechanism is designed to prevent model collapse but inherently limits the performance of S-DPO to the capabilities of the reference model. Given the inherent limitations of LML, the suboptimal performance of the reference model ultimately undermines the overall effectiveness of S-DPO (\cf Section \ref{sec:cp_exp}).
\item \textbf{Unstable Performance.} S-DPO relies on sampling negative items to establish the ranking relationship between positive and negative items. However, this sampling process incurs performance instability. To evaluate this, we train the model using multiple random seeds and calculated the deviation of NDCG@10. 
% As shown in Table \ref{tab:deviation}, the deviation of S-DPO is significantly higher than that of our proposed MSL, highlighting the instability of S-DPO's performance.
As shown in Table \ref{tab:deviation}, S-DPO exhibits significantly higher deviation compared to our proposed MSL, with performance losses reaching up to 16.4\% on the Toy dataset and 15.3\% on the Book dataset. This highlights the instability of S-DPO's performance.
\item \textbf{High Computational Cost.} S-DPO requires further fine-tuning on the reference model, which entails additional training epochs. Furthermore, the inclusion of extra negative items substantially increases the data size. These factors contribute to its inefficiency. Empirically, S-DPO requires nearly four times the runtime of MSL and LML (\cf Section \ref{efficiency}).

\end{itemize}

\section{Methodology}
In this section, we first detail the proposed Masked Softmax Loss (MSL) to address the limitations of language modeling loss (Subsection \ref{sec:MSL}). We then highlight the potential gradient vanishing challenge in MSL and propose the Adaptive Temperature Strategy to tackle this issue (Subsection \ref{sec:ATS}). The schematic diagrams of the MSL and LML methods are shown in Figure \ref{method}.

\subsection{Masked Softmax Loss}
\label{sec:MSL}
The above analyses reveal that the limitations of LML primarily lie in the penalization of invalid tokens --- it not only causes the loss to deviate from the recommendation objective but also introduces improper negative signals. To address this, a straightforward approach is to mask the invalid tokens in LML, \ie directly leverage the second component of LML to optimize LLMs. Formally, the Masked Softmax Loss is formulated as follows:

% \begin{equation}
%     \mathcal{L}_{mask}(x,y^p;\theta) = \sum_{t=1}^{|y^p|} - \log \frac{\exp(f_\theta(y_t^p|x,y^p_{<t})))}{\sum_{z \in \mathcal{Z}_{valid}(y^p_{<t})} \exp(f_\theta(z|x,y^p_{<t})))}
% \end{equation}
\begin{equation}
    \mathcal{L}_{MSL}(x,y^p;\theta) = \sum_{t=1}^{|y^p|} - \log \frac{\exp(f_\theta(y_t^p))}{\sum_{z \in \mathcal{Z}_{valid}} \exp(f_\theta(z))}
\end{equation}

This simple strategy effectively addresses the limitations by eliminating the penalization of invalid tokens. 
One might be concerned that this strategy could increase the risk of hallucination \cite{ji2023towards}, where LLMs generate fictitious item descriptions during the inference stage. This concern can be easily mitigated by employing constrained beam search during generation \cite{de2020autoregressive}. Specifically, when choosing or sampling the next token in beam search, the selection can be restricted to valid tokens rather than the entire vocabulary. 
% i.e., based on
% \begin{equation}
%     P^{'}_\theta(z|x,r_{<t}) = 
%     \begin{cases} 
%         P_\theta(z|x,r_{<t}), & z \in \mathcal{Z}_{valid}(r_{<t}) \\ 
%         0, & z \notin \mathcal{Z}_{valid}(r_{<t})
%     \end{cases} 
% \end{equation}
Such strategy ensures that the generated content corresponds to a valid item in the system, effectively mitigating the hallucination issue.

Overall, MSL possesses the following desirable properties:

\textbf{Alignment with the Recommendation Objective.} Intuitively, masking invalid tokens guide the model to focus more on differentiating positive items from negative ones. In fact, we have the following lemma establishing the theoretical connections between MSL and NDCG:
% \begin{lemma}
% \label{lemma_ndcg}
% Considering a LLM-based RS that leverages the scores $P_\theta^{valid}(y^v|x)$ for ranking items, where \( P_\theta^{valid}(y^v|x) \) represents the probability of item \( v \) within the valid item set \( \mathcal{V} \). Optimizing $\mathcal{L}_{mask}(x,y^p;\theta)$ serves as a tighter upper bound of $-log NDCG(S)$ compared with $\mathcal{L}(x,y^p;\theta)$, \ie $-log NDCG(S) \leq \mathcal{L}_{mask}(x,y^p;\theta) \leq \mathcal{L}(x,y^p;\theta)$. 
% \end{lemma}
\begin{lemma}
\label{lemma_ndcg}
Considering a LLM-based RS that leverages the scores $P_\theta^{valid}(y^v|x) = \prod_{t=1}^{|y^v|} P_\theta^{valid}(y^v_t|x,y^v_{<t})$ for ranking items, where
$$P_\theta^{valid}(y^v_t|x,y^v_{<t}) = \frac{\exp(f_\theta(y_t^v))}{\sum_{z \in \mathcal{Z}_{valid}(y^v_{<t})} \exp{(f_\theta(z))}}$$ 
represents the probability of the token \( y^v_t \) within the valid token set \( \mathcal{Z}_{valid}(y^v_{<t}) \), optimizing $\mathcal{L}_{MSL}(x,y^p;\theta)$ serves as a tighter upper bound of $-log NDCG(S)$ compared with $\mathcal{L}_{LML}(x,y^p;\theta)$, \ie 
$$-log NDCG(S) \leq \mathcal{L}_{MSL}(x,y^p;\theta) \leq \mathcal{L}_{LML}(x,y^p;\theta)$$ 
\end{lemma}
The proof is presented in appendix \ref{apd:lemma_ndcg}. Note that the premise of ranking items based on $P_\theta^{valid}(y^v|x)$ is naturally satisfied when we mask invalid tokens during generation using constrained beam search with a large beam size. This lemma demonstrates that MSL is well-aligned with the recommendation objective and provides a tighter upper bound for optimizing NDCG compared to LML. \(\mathcal{L}_{LML}^1\) in LML is redundant for NDCG optimization and may even introduce interference as previously discussed. Consequently, MSL is theoretically anticipated to achieve superior performance.

\textbf{Ease of Implementation.} Our MSL is simple, easily implemented, and can serve as a suitable surrogate for LML with minimal code revisions. The main implementation complexity lies in identifying valid tokens. 
In fact, this can be easily achieved by using a trie tree (\aka a prefix tree) \cite{bodon2003trie}. We can utilize existing packages of \textit{marisa-trie} (with only 3 lines of codes) to construct the trie tree from all item descriptions during the pre-processing stage and calculate the masking matrix. Subsequently, we can revise the LML to MSL by simply applying the masking matrix. MSL can be seamlessly integrated into various existing LLM-based recommendation methods, including the recently proposed BIGRec \cite{bao2023bi}, LLaRA \cite{liao2024llara}, A-LLM \cite{kim2024large}, and consistently yield improvements (\cf section \ref{sec:cp_exp}).

% \textbf{Efficiency.} Regarding efficiency, the primary complexity is in constructing the trie tree and the masking matrix. However, this stage is highly efficient with a complexity of $O(|I|\hat l)$ in both time and memory, where $|I|$ denotes the number of items on the system, and $\hat l$ denotes the max number of tokens for item descriptions.  Thus, our MSL is often more efficient than the original language loss.

\textbf{Efficiency.} The primary computational challenge lies in constructing the trie tree and the masking matrix. However, this process is highly efficient, with a time and memory complexity of \(O(|\mathcal{V}|\hat{l})\), where \(|\mathcal{V}|\) denotes the number of items in the system, and \(\hat{l}\) represents the average token length of the item description. Empirically, the Trie tree construction for all datasets is completed in under one second. Furthermore, MSL improves efficiency by excluding invalid tokens from the loss calculation (\cf section \ref{efficiency}).

\subsection{Adaptive Temperature Strategy}
\label{sec:ATS}

\subsubsection{Potential Gradient Vanishing Issue}
Despite the theoretical advantages of MSL, it may encounter gradient vanishing issues in practical applications, challenging its effectiveness. To illustrate this effect, the gradient of MSL over each sample can be expressed as follows:
\begin{equation}
\begin{aligned}
\nabla_\theta \mathcal{L}_{MSL}(x, y^p; \theta) &= -\sum_{t=1}^{|y^p|} w(y_t^p) g(y_t^p,\theta)
\end{aligned}
\end{equation}
where 
\begin{align}
% w(y_t^p) &= 1 - P^{valid}_{\theta}(y_t^p) = 1 - \frac{\exp(f_\theta(y_t^p)))}{\sum_{z \in \mathcal{Z}_{valid}} \exp{(f_\theta(z))}} \\
% &w(y_t^p) = 1 - P^{valid}_{\theta}(y_t^p|x,y_{<t}^p) \\
&w(y_t^p) = 1 - P^{valid}_{\theta}(y_t^p\mid x, y^p_{<t}) \\
% &P^{valid}_{\theta}(y_t^p|x,y_{<t}^p) = \frac{\exp(f_\theta(y_t^p)))}{\sum_{z \in \mathcal{Z}_{valid}} \exp{(f_\theta(z))}} \\
% P_\theta^{valid}(y_t^p) &= \frac{\exp(f_\theta(y_t^p)))}{\sum_{z \in \mathcal{Z}(y^p_{<t})} \exp{(f_\theta(z))}} \\
&g(y_t^p,\theta) = \nabla_\theta f_\theta(y_t^p) - \frac{\sum_{z \in \mathcal{Z}'} \exp(f_\theta(z)) \nabla_\theta f_\theta(z)}{\sum_{z \in \mathcal{Z}'} \exp(f_\theta(z))} 
% g(x,y_t^p,\theta) = \left(\nabla_\theta f_\theta(y_t^p) - \sum_{z \in \mathcal{Z}_{valid} \setminus \{y_t^p\}} P_\theta^{''} \nabla_\theta f_\theta(z)\right)
\end{align}
% $P_\theta^{valid}(y_t^p|x,y_{<t}^p)$ represents the probability of \(y_t^p\) within the set of valid tokens \(\mathcal{Z}_{valid}\). 
$\mathcal{Z}' = \mathcal{Z}_{valid} \setminus \{y_t^p\}$ represents the set of negative tokens. For simplicity, let \(P_{\theta}^{valid}(y^p_t)\) represent \(P_{\theta}^{valid}(y^p_t \mid x, y^p_{<t})\).

\begin{figure}[t]
  \centering
  \includegraphics[width=\linewidth]{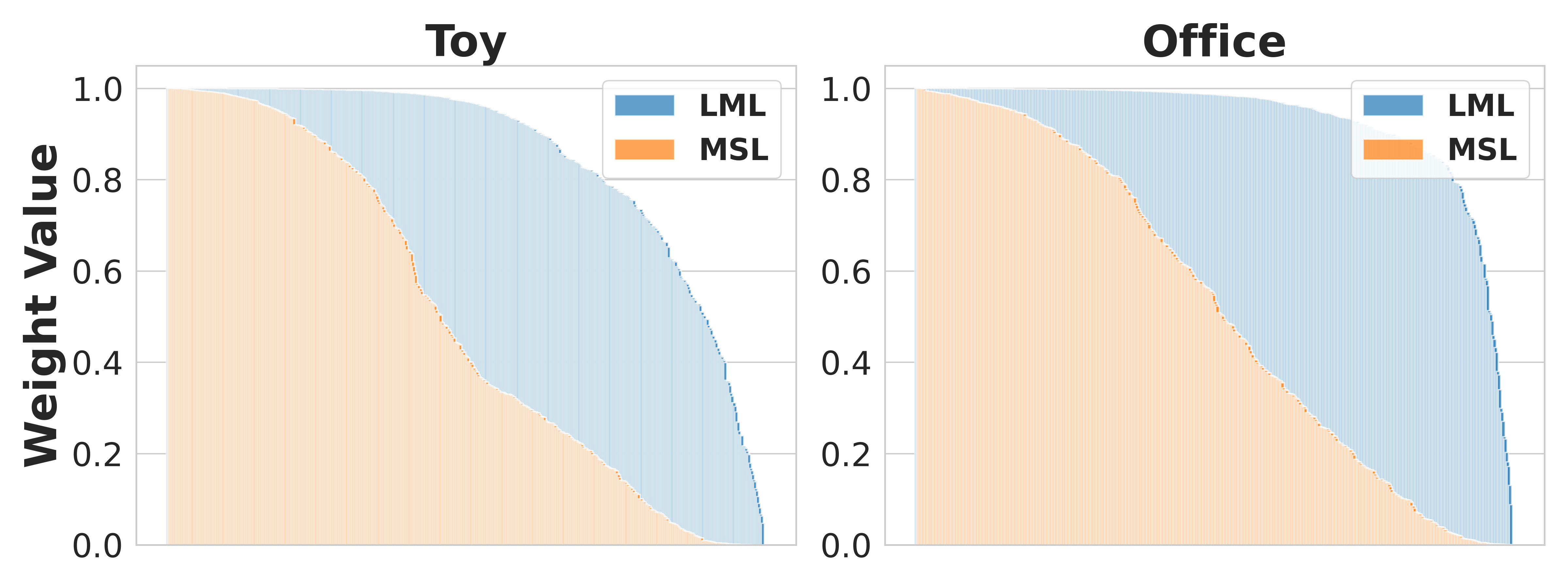}
  \caption{The weight value $w(y_t^p)$ distribution of a batch of samples for MSL and LML. The weight are sorted in descending order.}
  \Description{}
  \label{LML_MSL_grad}
  % \vspace{-0.6cm}
\end{figure}

As observed, the magnitude of gradient is influenced by the weight $w(y_t^p)$. The gradient vanishing phenomenon can be attributed to a reduced number of terms in the denominator of \(P_\theta^{valid}(y_t^p)\) as compared with \(P_\theta(y_t^p)\), which naturally increases \(P_\theta^{valid}(y_t^p)\), thereby decreasing the weight \(w(y_t^p)\). This reduction can even cause the gradient to approach zero, particularly because the logits of positive tokens are often larger than those of other valid tokens\footnote{This assumption is reasonable, as the optimization process tends to increase the logits of positive tokens while decreasing those of negative tokens.}.

Figure \ref{LML_MSL_grad} presents the weight values of a batch of samples for both LML and MSL. When the valid token mask is applied to MSL, the weight values of all samples are substantially reduced, with some values even nearing zero. 
Importantly, these tokens are often crucial, as they are typically located among the first few tokens in the response and play a pivotal role in training. 
Empirical analysis on the Office dataset underscores this point: 61\% of samples with weight values below 0.1 are concentrated within the first three tokens of the item. Similar patterns are observed across other datasets.

\subsubsection{The Introduction of Temperature}
To tackle this issue, we have found that the introduction of a temperature $\tau$ can effectively address this problem:
\begin{equation}
    \mathcal{L}_{MSL}(x,y^p;\theta) = \sum_{t=1}^{|y^p|} - \tau \log \frac{\exp(f_\theta(y_t^p) / \tau)}{\sum_{z \in \mathcal{Z}_{valid}} \exp(f_\theta(z) / \tau)}
\end{equation}
where the weight $w(y_t^p)$ can be written as: 
\begin{equation}
w(y_t^p) = 1 - P^{valid}_{\theta}(y_t^p) = 1 - \frac{\exp(f_\theta(y_t^p) / \tau)}{\sum_{z \in \mathcal{Z}_{valid}} \exp{(f_\theta(z) / \tau)}}
\end{equation}

\begin{figure}[t]
  \centering
  \includegraphics[width=\linewidth]{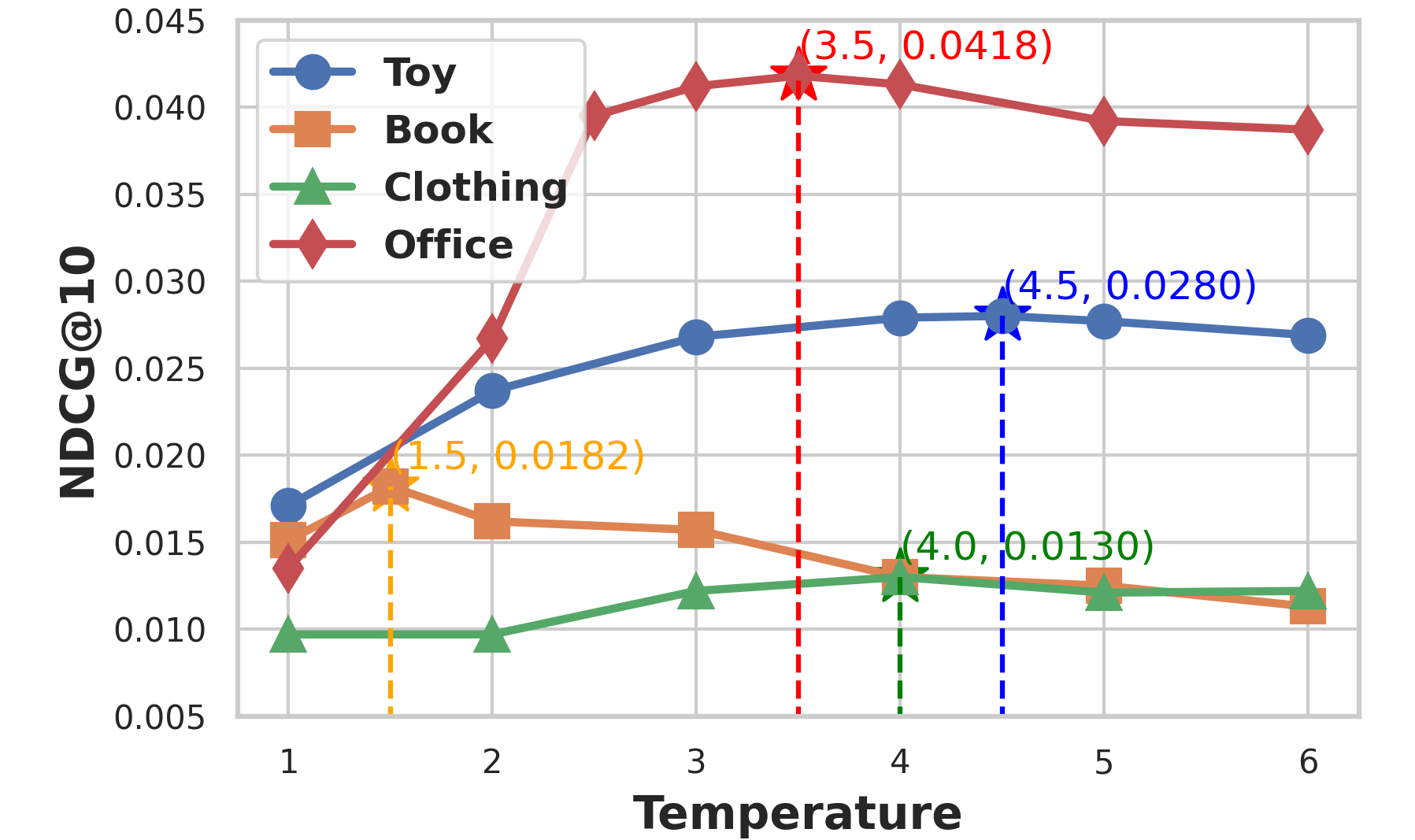}
  \caption{The performance of MSL across different temperature $\tau$.}
  \Description{}
  \label{ndcg_temperature}
  % \vspace{-0.6cm}
\end{figure}

The introduction of temperature can modulate the magnitude of the gradient. Considering that the logits of positive tokens are typically larger than those of other tokens, an increase in $\tau$ would relatively reduce the value of $P^{valid}_{\theta}(y_t^p)$, increasing \(w(y_t^p)\); 
% conversely, a decrease in $\tau$ would amplify the logits differences between the positive and other tokens, thus increasing $P_{valid}(y_t^p|x,y^p_{<t})$ and decreasing w. 
% Figure \ref{ndcg_temperature} shows the effectiveness of introducing temperature, which significantly boosts performance. In contrast, strategies like tuning learning-rate to amplify the effect of valid tokens do not yield satisfactory results. 
Figure \ref{ndcg_temperature} highlights the impact of incorporating temperature, which leads to a significant improvement in performance. Conversely, alternative approaches, such as adjusting the learning rate or introducing a balancing coefficient for negative tokens, fail to yield satisfactory results. The empirical evidence supporting these findings will be presented in Section \ref{sec:ablation}.

\subsubsection{Adaptive Temperature Strategy}
Despite its effectiveness, the introduction of $\tau$ incurs another hyperparameter tuning challenge. Given that the average number of valid tokens varies across different datasets, the optimal value of $\tau$ naturally evolves. For example, the optimal $\tau$ on dataset Book is 1.5, while it is 4.5 on Toy as shown in Figure \ref{ndcg_temperature}. Transferring the optimal $\tau$ from one dataset to another without adjustment can lead to significant performance drops. This necessitates extensive hyperparameter tuning of $\tau$, which can be particularly time-consuming, especially for heavy LLM-based recommenders.

To address this, inspired by recent studies on temperature \cite{chen2023adap}, we develop an Adaptive Temperature Strategy (ATS) for MSL. This strategy dynamically and adaptively adjusts $\tau$ to ensure that $P_\theta^{valid}(y_t^p)$ remains close to a target value $\eta$, preventing it from becoming excessively large and incurring gradient vanishing. Specifically, we have the following lemma:
\begin{lemma}
\label{lemma_ada_tau}
For each training token instance $(x, y_t^p)$, assuming the logits of the valid tokens $f_\theta(z), z \in \mathcal{Z}_{valid}$ follow a Gaussian distribution $\mathcal{N}(\mu_t, \sigma_t^2)$. Then $\tau_t$ for the equation $P_\theta^{valid}(y_t^p) = \eta$ can be approximated as:
\begin{equation}
\label{ada_tau}
\tau_t \approx \frac{(f_\theta(y_t^p) - \mu_t)-\sqrt{(f_\theta(y_t^p) - \mu_t)^2-2 \sigma^2_t \log (|\mathcal{Z}_{valid}| \eta )}}{2\log (|\mathcal{Z}_{valid}| \eta)}
\end{equation}
\end{lemma}

% \begin{lemma}
% Let $\mathbb{L}_t$ be the distribution of $\mathbf{l}^t$ over all instances. Let $\mathbf{l}^t$ be a random variable that sampled from $\mathbb{L}_t$. Suppose the distribution $\mathbb{L}_t$ have a sub-exponential tail such that the following conditions hold for some $\lambda>0$ :

% $$
% \begin{aligned}
% & p\left(\left(\mathbf{l}^t-\mathbb{E}_{\mathbb{L}_t}[\mathbf{l}^t]\right)>b\right) \leq 2 e^{-2 b / \lambda}
% \end{aligned}
% $$

% When $\tau_t \geqslant \max \left(2 \lambda, T\right)$, where $T$ is the lower bound of $\tau_t$ it can be approximated as:

% $$
% \tau_t \approx \frac{(l_{y_t}-\mu_t)-\sqrt{(l_{y_t}-\mu_t)^2-2 \sigma^2_t \log m_t \eta }}{2\log m_t \eta}
% $$
% where \(m_t\) represents the number of elements in the set of valid tokens \(\mathcal{T}_{\text{valid}}(t)\). \(\mu_t\) and \(\sigma^2_t\) denote the mean and variance of \(\mathbf{l}^t\), respectively.
% \end{lemma}

The proof is presented in the appendix \ref{apd:lemma_adatau}. The proof references the work \cite{chen2023adap} but adapts the process to token-wise LLM-based recommendation scenarios and different distribution conditions. The assumption of a Gaussian distribution nearly holds, as discussed in the appendix \ref{apd:gaussian}.

Eq.(\ref{ada_tau}) gives the token-wise optimal configuration of $\tau_t$. To make the training more stable and reduce the extra effort of calculating the token-wise $\tau_t$, we prefer to set a global uniform $\tau$ across various training instances:
\begin{equation}
\label{ada_tau_2}
\tau \approx \frac{(f_\theta(y_t^p) - \mu)-\sqrt{(f_\theta(y_t^p) - \mu)^2-2 \sigma^2 \log m \eta }}{2\log m \eta}
\end{equation}
where $\mu$, $\sigma^2$ denote the mean and variance of $f_\theta(z)$ for all training instances, and $m$ denotes the average number of valid tokens. This Eq.(\ref{ada_tau_2}) can adaptively adjust the value of $\tau$ according to the current model state and the average number of valid tokens in the datasets, serving as an efficient alternative to brute-force hyperparameter search.

\begin{table}[]
\centering
\caption{Statistics of the datasets. AVT represents the average number of valid tokens per token instance.}
\label{tab:statistics}
{%
\begin{tabular}{@{}lccccc@{}}
\toprule
Dataset  & \#Users & \#Items & \#Interactions & \#Density & \#AVT  \\ \midrule
Toy      & 19124  & 11758  & 165247        & 0.0735\%  & 54.38 \\
Book     & 16559  & 6344   & 151928        & 0.1446\%  & 70.90 \\
Clothing & 39230  & 22948  & 277534        & 0.0308\%  & 53.74 \\
Office   & 4895   & 2414   & 53149         & 0.4498\%  & 8.98 \\ \bottomrule
\end{tabular}%
}
\end{table}

\section{Experiments}
We aim to answer the following research questions:
\begin{itemize}[left=5pt]
\item \textbf{RQ1:} How does MSL perform compare to existing state-of-the-art recommendation methods?
\item \textbf{RQ2:} How do different components of MSL affect?
\item \textbf{RQ3:} How does MSL perform compared with state-of-the-art in terms of both accuracy and efficiency?
\end{itemize}

% Please add the following required packages to your document preamble:
% \usepackage{booktabs}
% \usepackage{multirow}
% \usepackage[table,xcdraw]{xcolor}
% Beamer presentation requires \usepackage{colortbl} instead of \usepackage[table,xcdraw]{xcolor}
\begin{table*}[]
\centering
\caption{The performance comparison on four real-world datasets. The best result is bolded. Improvement denotes the improvement of MSL over the best results obtained using LML and S-DPO. "N" represents NDCG, and "H" represents Hit Ratio.}
\label{tab:performance_cp}
\resizebox{\textwidth}{!}{%
\begin{tabular}{@{}l|cccc|cccc|cccc|cccc@{}}
\toprule
                         & \multicolumn{4}{c|}{Toy}                                              & \multicolumn{4}{c|}{Book}                                             & \multicolumn{4}{c|}{Clothing}                                         & \multicolumn{4}{c}{Office}                                            \\ \cmidrule(l){2-17} 
\multirow{-2}{*}{Method} & N@5             & N@10            & H@5             & H@10            & N@5             & N@10            & H@5             & H@10            & N@5             & N@10            & H@5             & H@10            & N@5             & N@10            & H@5             & H@10            \\ \midrule
SASRec                   & 0.0101          & 0.0126          & 0.0190          & 0.0265          & 0.0097          & 0.0133          & 0.0176          & 0.0285          & 0.0046          & 0.0056          & 0.0086          & 0.0116          & 0.0132          & 0.0183          & 0.0260          & 0.0421          \\
BERT4Rec                 & 0.0157          & 0.0191          & 0.0229          & 0.0336          & 0.0118          & 0.0171          & 0.0187          & 0.0351          & 0.0071          & 0.0093          & 0.0110          & 0.0180          & 0.0225          & 0.0307          & 0.0358          & 0.0618          \\
SASRec+DROS              & 0.0129          & 0.0160          & 0.0217          & 0.0311          & 0.0110          & 0.0156          & 0.0196          & 0.0340          & 0.0050          & 0.0067          & 0.0088          & 0.0142          & 0.0130          & 0.0226          & 0.0260          & 0.0561          \\ \midrule
LLM-CF                   & 0.0103          & 0.0132          & 0.0186          & 0.0275          & 0.0106          & 0.0142          & 0.0178          & 0.0292          & 0.0041          & 0.0052          & 0.0074          & 0.0116          & 0.0144          & 0.0192          & 0.0239          & 0.0395          \\
DLLM4Rec                 & 0.0104          & 0.0134          & 0.0190          & 0.0284          & 0.0099          & 0.0136          & 0.0178          & 0.0303          & 0.0042          & 0.0061          & 0.0082          & 0.0138          & 0.0137          & 0.0198          & 0.0239          & 0.0504          \\ \midrule
BIGRec+LML               & 0.0138          & 0.0182          & 0.0213          & 0.0353          & 0.0109          & 0.0137          & 0.0169          & 0.0258          & 0.0047          & 0.0073          & 0.0092          & 0.0174          & 0.0113          & 0.0220          & 0.0203          & 0.0530          \\
BIGRec+SDPO              & 0.0174          & 0.0219          & 0.0271          & 0.0413          & 0.0118          & 0.0145          & 0.0189          & 0.0276          & 0.0062          & 0.0089          & 0.0114          & 0.0198          & 0.0129          & 0.0256          & 0.0239          & 0.0629          \\ \midrule
\rowcolor[HTML]{C0F1FF} 
BIGRec+MSL               & 0.0245          & 0.0288          & 0.0357          & 0.0488          & 0.0125          & 0.0172          & 0.0214          & 0.0356          & 0.0091          & 0.0120          & 0.0146          & 0.0236          & \textbf{0.0402} & \textbf{0.0438} & \textbf{0.0556} & 0.0665          \\ \midrule
\rowcolor[HTML]{C0F1FF} 
\textit{Improvement}     & 41.06\%          & 31.35\%          & 31.54\%          & 18.18\%          & 5.70\%           & 18.55\%          & 13.25\%          & 28.93\%          & 46.05\%          & 34.52\%          & 28.07\%          & 19.19\%          & 213.17\%         & 70.80\%          & 132.61\%         & 5.79\%           \\ \midrule
LLaRA+LML                & 0.0145          & 0.0193          & 0.0225          & 0.0375          & 0.0102          & 0.0132          & 0.0173          & 0.0265          & 0.0050          & 0.0083          & 0.0088          & 0.0190          & 0.0093          & 0.0204          & 0.0177          & 0.0514          \\
LLaRA+SDPO               & 0.0164          & 0.0211          & 0.0250          & 0.0394          & 0.0094          & 0.0114          & 0.0128          & 0.0187          & 0.0077          & 0.0099          & 0.0138          & 0.0206          & 0.0196          & 0.0242          & 0.0379          & 0.0524          \\ \midrule
\rowcolor[HTML]{C0F1FF} 
LLaRA+MSL                & 0.0233          & 0.0283          & 0.0336          & 0.0488          & \textbf{0.0135} & 0.0175          & \textbf{0.0244} & 0.0365          & \textbf{0.0108} & \textbf{0.0140} & \textbf{0.0176} & \textbf{0.0274} & 0.0374          & 0.0417          & 0.0535          & \textbf{0.0670} \\ \midrule
\rowcolor[HTML]{C0F1FF} 
\textit{Improvement}     & 41.98\%          & 34.05\%          & 34.17\%          & 23.81\%          & 31.96\%          & 32.71\%          & 40.79\%          & 37.93\%          & 40.12\%          & 41.61\%          & 27.54\%          & 33.01\%          & 90.87\%          & 72.08\%          & 41.10\%          & 27.82\%          \\ \midrule
A-LLM+LML                & 0.0151          & 0.0197          & 0.0236          & 0.0378          & 0.0107          & 0.0132          & 0.0173          & 0.0251          & 0.0055          & 0.0082          & 0.0100          & 0.0182          & 0.0110          & 0.0219          & 0.0192          & 0.0519          \\
A-LLM+SDPO               & 0.0155          & 0.0201          & 0.0248          & 0.0388          & 0.0065          & 0.0093          & 0.0094          & 0.0180          & 0.0072          & 0.0104          & 0.0116          & 0.0216          & 0.0189          & 0.0240          & 0.0306          & 0.0468          \\ \midrule
\rowcolor[HTML]{C0F1FF} 
A-LLM+MSL                & \textbf{0.0248} & \textbf{0.0296} & \textbf{0.0365} & \textbf{0.0513} & 0.0130          & \textbf{0.0184} & 0.0228          & \textbf{0.0395} & 0.0097          & 0.0127          & 0.0158          & 0.0252          & 0.0353          & 0.0388          & 0.0488          & 0.0597          \\ \midrule
\rowcolor[HTML]{C0F1FF} 
\textit{Improvement}     & 59.81\%          & 47.73\%          & 47.06\%          & 32.26\%          & 21.01\%          & 39.44\%          & 31.58\%          & 57.27\%          & 34.65\%          & 22.54\%          & 36.21\%          & 16.67\%          & 86.84\%          & 61.46\%          & 59.32\%          & 15.00\%          \\ \bottomrule
\end{tabular}%
}
\end{table*}

\subsection{Experimental Settings}
\subsubsection{Datasets}
Four conventional real-world datasets: \textit{Amazon Toys and Games}, \textit{Amazon Books}, \textit{Amazon Clothing, Shoes and Jewelry} and \textit{Amazon Office Products} \footnote{\url{https://jmcauley.ucsd.edu/data/amazon/index_2014.html}} are utilized in our experiments, which are commonly used for the studies of LLM-based recommendation \cite{cui2024distillation, bao2023bi, cao2024aligning, lee2024star, li2023e4srec}. To ensure a fair comparison, we adopt the same data preprocessing used in recent studies \cite{bao2023bi, cui2024distillation}. Specifically, we firstly apply the 5-core setting to the original dataset, then for user interaction sequences longer than 11 interactions, a sliding window of length 11 is applied to segment the sequences. The resulting sequences are then sorted in ascending order by timestamp and split into training, validation, and testing sets with an 8:1:1 ratio. We randomly retain 100,000 items for \textit{Amazon Books} before 5-core processing due to its large size. The processed dataset statistics are presented in Table \ref{tab:statistics}.

\subsubsection{Baselines}
The methods compared fall into several categories:
\begin{itemize}[left=5pt]
\item \textbf{Traditional recommenders (SASRec \cite{kang2018self}, BERT4Rec \cite{sun2019bert4rec}, DROS \cite{yang2023generic})} SASRec utilizes a self-attention-based model to capture user interests. BERTRec adopts the BERT to bidirectional model user preferences. DROS incorporates DRO to improve the model's resilience to distributional shifts.
\item \textbf{LLM-enhanced recommenders (DLLM2Rec \cite{cui2024distillation}, LLM-CF \cite{sun2024large})} DLLM2Rec introduces a distillation module designed to bridge the performance gap between LLMs and traditional RS. LLM-CF enhances traditional RS by integrating reasoning-driven collaborative filtering features derived from LLMs using CoT techniques. We use SASRec as the backbone for LLM-enhanced recommenders.
\item \textbf{LLM-based recommenders (BIGRec \cite{bao2023bi}, LLaRA \cite{liao2024llara}, A-LLM \cite{kim2024large})} BIGRec develops instruction-tuning templates to fine-tuning LLMs on RS datasets. LLaRA enhances collaborative signals by incorporating embeddings produced by traditional models into prompts. A-LLM further alignes these embeddings with corresponding textual information.
\item \textbf{Improved Loss Function for LLM-based Recommenders (S-DPO \cite{chen2024softmax})} S-DPO leverages the DPO to guide LLMs using the ranking information of positive and negative samples.
\end{itemize}

\subsubsection{Implementation Details} 
LLaMA3 8B model \cite{dubey2024llama} is utilized as the backbone of all the LLM-based recommenders.
As for training LLM-based recommenders, we train the models for 10 epochs and report the results of the epoch with the highest NDCG@5. 
% During inference, we adopt constrained beam search to mitigate hallucination and prevent the generation of fictitious items as suggested in prior work \cite{zhu2024collaborative}. 
For inference, we evaluate two mainstream methods as baselines: grounding \cite{bao2023bi} and constrained beam search \cite{zhu2024collaborative}, and we report the better-performing results. The ranking results obtained from constrained beam search are used to construct the recommendation list, with the number of beams fixed at 10.
For MSL, we only modify the loss function of the original backbone while following its original hyperparameter settings. The parameter \(\eta\) is set to 0.25.
% For both traditional and LLM-enhanced recommenders, we fix the learning rate at 0.001 and the embedding size at 64. 
To ensure fair comparisons, we leverage the source code provided in the original papers and tune the hyperparameters of all baseline methods following the guidelines specified in their respective works. 
% The random seed is fixed to 42 in all experiments.
Two widely-used metrics \textit{NDCG@K} and \textit{Hit Ratio@K} are employed for evaluating the recommendation accuracy (\textit{K} = 5, 10).

\subsection{Performance Comparison (RQ1)}
\label{sec:cp_exp}
Table \ref{tab:performance_cp} provides a comparative analysis of the performance of the proposed MSL method against baseline approaches. 

\textbf{MSL demonstrates a significant enhancement in the performance of various LLM-based recommenders.} MSL consistently outperforms all baseline across all datasets. This remarkable improvement can be attributed to the design of MSL as a specialized loss function tailored for LLM-based recommenders. 
% MSL substantially boosts performance by avoiding ineffective optimization and enabling LLMs to focus more effectively on recommendation-specific objectives.

\textbf{The performance improvements of LLM-enhanced recommenders remain relatively limited.} LLM-CF exhibits negative gains on three out of the four datasets. This underperformance is primarily due to the inherent gap between LLMs and traditional models, which hinders the effective transfer of knowledge. 
% Besides, the semantic embeddings generated by LLMs may even introduce noise, adversely affecting the model performance. 
DLLM2Rec, which directly generates ranking results using LLMs and incorporates a specially designed distillation mechanism, partially addresses this gap. However, its performance remains constrained by the limitations of LLMs' recommendation capabilities as teacher models.

\textbf{S-DPO demonstrates limited and inconsistent performance improvements in LLM-based recommendation systems.} Its dependence on the reference model, coupled with instability introduced by sampling, significantly constrains its effectiveness. As a result, S-DPO often fails to deliver consistent improvements, and in some cases, even demonstrates negative performance gains. For example, NDCG@10 of A-LLM+S-DPO decreases from 0.0132 to 0.0093 on the Book dataset.

% \textbf{S-DPO shows limited and unstable performance improvements for LLM-based recommenders.} 
% S-DPO incorporates the ranking relationships between positive and negative samples into the model training process based on DPO. However, due to the constraints imposed by DPO to minimize the gap between the model and the reference model, the performance of S-DPO is limited by the capabilities of the reference model, resulting in relatively modest performance improvements.
% Furthermore, the use of a small number of negative samples (set as 3 in the original implementation) introduces the risk of erroneous ranking information, leading to unstable results (\cf Section \ref{sec:analyse}). For example, on the Book dataset, the NDCG@10 of A-LLM+S-DPO decreases from 0.0132 to 0.0093 compared to A-LLM. Additionally, S-DPO requires additional sampling of negative samples, which causes the data size to grow linearly, thereby reducing computational efficiency. For a detailed analysis of efficiency comparisons, refer to Section \ref{efficiency}.

\subsection{Ablation Study (RQ2)}
\label{sec:ablation}
Table \ref{tab:ablation_study} presents the results of ablation study. Specifically, we investigate the effects of MSL, temperature $\tau$ and the ATS module, as well as alternative approaches to mitigating the vanishing gradient problem.

\textbf{Adjusting the temperature $\tau$ significantly enhances MSL performance.} MSL without temperature underperforms compared to LML on certain datasets (e.g., Toy and Office) due to the vanishing gradient issue. Employing an hyperparameter search strategy for the $\tau$ effectively addresses this issue, resulting in substantial performance improvements. This highlights the critical role of $\tau$ in unlocking MSL's potential. Notably, a similar hyperparameter search for LML revealed only marginal performance gains, suggesting that the observed improvements are attributable to MSL itself rather than the tuning of $\tau$. 

\textbf{The ATS module rivals or surpasses brute-force hyperparameter search in effectiveness.} This demonstrates the efficacy of ATS as a dynamic optimization strategy. ATS offers better flexibility than fixed temperature and eliminates the need for exhaustive manual tuning.

\textbf{Alternative strategies for mitigating vanishing gradients show limited effectiveness.} We also investigate two alternative strategies to address the vanishing gradient problem in MSL: adjusting the learning rate (i.e., MSL + tuning lr) and introducing a balancing coefficient \(\alpha\) for negative tokens (i.e., MSL + \(\alpha\), where the negative token component in the denominator of \(\mathcal{L}_{MSL}\) is multiplied by \(\alpha\)). 
\(\alpha\) is set as \(|\mathcal{Z}| / |\mathcal{Z}_{\text{valid}}|\).
While these adjustments yielded minor improvements, the results remained significantly inferior to those achieved through temperature adjustment. 
% Additionally, we attempted to introduce a balancing coefficient $\alpha$ for negative tokens in an effort to amplify the gradients based on the number of valid tokens. The formal definition is as follows:
% \begin{equation}
%      \sum_{t=1}^{|y^p|} - \log \frac{\exp(f_\theta(y_t^p)))}{\exp(f_\theta(y_t^p)) + \alpha \sum_{z \in \mathcal{Z}_{valid} \setminus \{y_t^p\}} \exp(f_\theta(z)))}
% \end{equation}
% where $\alpha = |\mathcal{Z}| / |\mathcal{Z}_{valid}|$. This strategy achieved certain improvements over MSL on the Toy and Book datasets, but resulted in performance degradation on the Clothing and Office datasets.
% These findings suggest that such strategies are less effective compared to temperature adjustment.

% Please add the following required packages to your document preamble:
% \usepackage{booktabs}
\begin{table}[]
\centering
\caption{Ablation study. Results are reported in NDCG@10. "+tuning \(\tau\)" indicates performing a hyperparameter search for the temperature \(\tau\). "+tuning lr" indicates performing a hyperparameter search for the learning rate. "+ \(\alpha\)" denotes introducing a coefficient for negative tokens. MSL (w/o \(\tau\)) represents MSL without the temperature. MSL (w/ ATS) represents MSL with the temperature adjusted using ATS.}
\label{tab:ablation_study}
\begin{tabular}{@{}l|cccc@{}}
\toprule
Method                & Toy    & Book   & Clothing & Office \\ \midrule
LML                   & 0.0182 & 0.0137 & 0.0073   & 0.0220 \\
LML + tuning   $\tau$   & 0.0182 & 0.0146 & 0.0074   & 0.0248 \\
MSL (w/o $\tau$)      & 0.0171 & 0.0151 & 0.0097   & 0.0135 \\
MSL + tuning lr         & 0.0184 & 0.0151 & 0.0101   & 0.0177 \\
MSL + $\alpha$ & 0.0202 & 0.0157 & 0.0078   & 0.0221 \\
MSL + tuning $\tau$   & 0.0280 & 0.0182 & 0.0130   & 0.0418 \\
MSL (w/ ATS)          & 0.0288 & 0.0172 & 0.0120   & 0.0438 \\ \bottomrule
\end{tabular}
\end{table}

% Please add the following required packages to your document preamble:
% \usepackage{booktabs}
% \usepackage{graphicx}
% \begin{table}[]
% \centering
% \caption{Average number of valid tokens for each dataset.}
% \label{tab:valid_token_num}
% \resizebox{\columnwidth}{!}{%
% \begin{tabular}{@{}lcccc@{}}
% \toprule
% Dataset                         & Toy   & Book  & Clothing & Office \\ \midrule
% Average number of valid tokens & 54.38 & 70.90 & 53.74    & 8.96   \\ \bottomrule
% \end{tabular}%
% }
% \end{table}

\subsection{Efficiency Comparison (RQ3)}
\label{efficiency}
\begin{figure}[t]
  \centering
  \includegraphics[width=\linewidth]{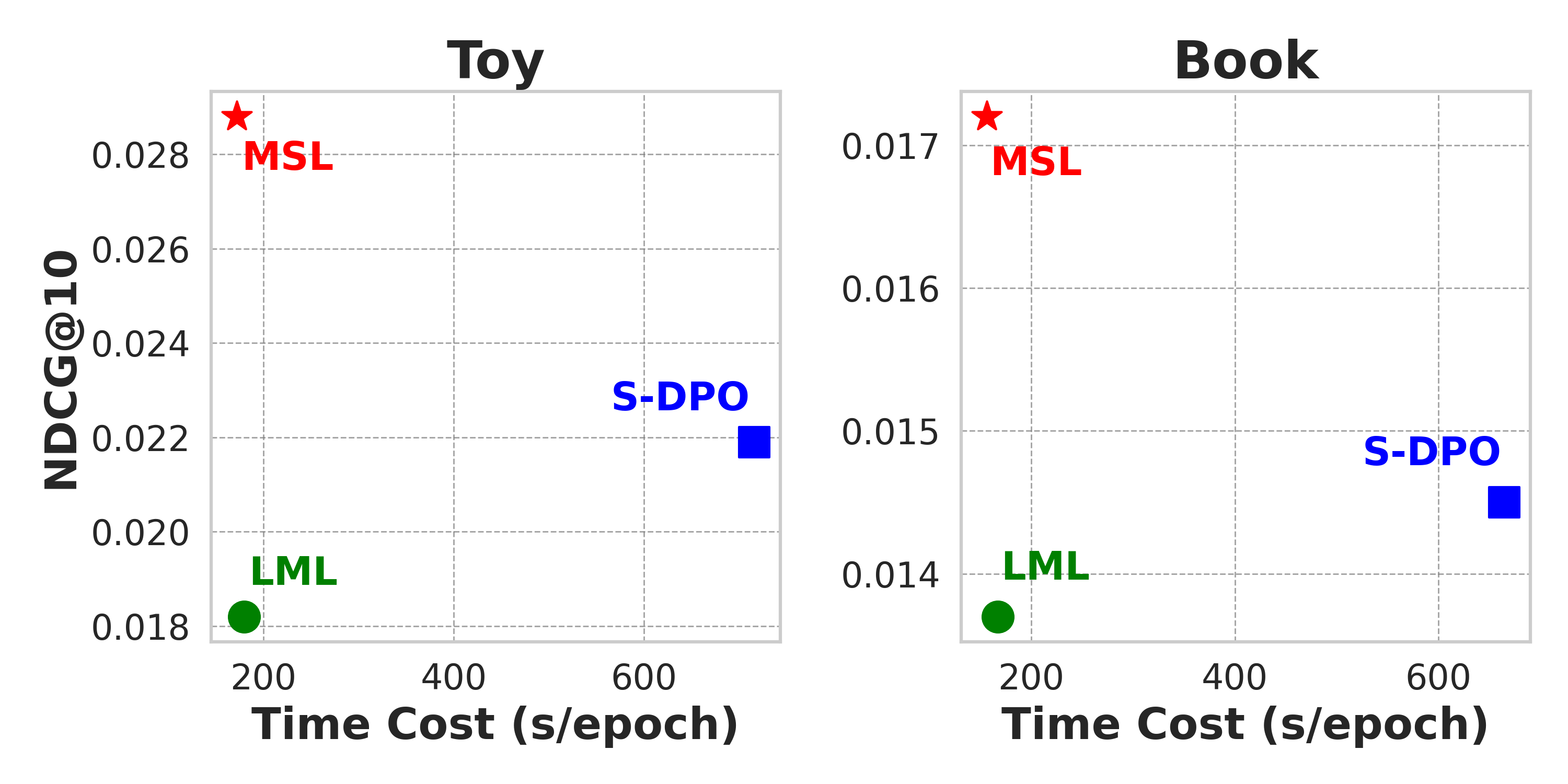}
  \caption{Performance comparisons in terms of both recommendation accuracy and efficiency.}
  \Description{}
  \label{efficiency_performance}
  % \vspace{-0.5cm}
\end{figure}

In this section, we analyze and compare the efficiency and performance of various methods. As illustrated in Figure \ref{efficiency_performance}, MSL achieves optimal recommendation performance while simultaneously demonstrating superior computational efficiency. Specifically, MSL improves efficiency by 315\% and 324\% on the Toy and Book datasets, respectively, compared to S-DPO. In comparison to LML, MSL achieves efficiency gains of 4.4\% and 6.7\% on the same datasets.

The enhanced efficiency of MSL stems from its ability to significantly reduce the number of invalid tokens. MSL restricts the scope to a small subset of valid tokens (the average number of valid tokens for each dataset is shown in Table \ref{tab:statistics}) compared to the 128,000 tokens in LLaMA3 vocabulary. This targeted approach minimizes computational overhead. 
The only additional overhead introduced by MSL arises from the construction of the Trie tree during the data preprocessing stage. As discussed in Section \ref{sec:MSL}, this process is highly efficient. Table \ref{tab:time_trie} shows that the consuming time for all datasets is consistently within one second.
% In contrast, S-DPO introduces additional computational overhead by requiring the sampling of \(n\) negative samples for each positive sample. This process increases the dataset size by a factor of \((n + 1)\), leading to a notable decline in runtime efficiency (approximately four times the runtime compared to MSL).
In contrast, S-DPO introduces additional computational overhead by requiring the sampling of \(n\) negative samples for each positive sample, leading to approximately four times runtime compared to MSL.

% Please add the following required packages to your document preamble:
% \usepackage{booktabs}
% \begin{table}[]
% \centering
% \caption{Comparison of the efficiency of different loss functions. The table reports the time required to run one epoch.}
% \label{tab:efficiency}
% \begin{tabular}{@{}lccc@{}}
% \toprule
% Dataset & SL   & S-DPO  & \textbf{MSL}  \\ \midrule
% Toy     & 166s & 1080s (+550.60\%) & \textbf{164s (-1.20\%)} \\
% Book    & 168s & 1083s (+544.64\%) & \textbf{156s (-7.14\%)} \\ \bottomrule
% \end{tabular}
% \end{table}

\section{Related Work}
\subsection{Sequential Recommendation}
Sequential recommendation aims to predict the next item of interest for a user based on their historical interactions. 
% Unlike traditional collaborative filtering, sequential recommendation explicitly considers the temporal order of interactions, enabling a more effective capture of the dynamic evolution of user preferences.
It has increasingly adopted various deep learning models over recent years. For example, GRU4Rec \cite{hidasi2015session} leverages recurrent neural networks (RNNs), while Caser \cite{tang2018personalized} utilizes convolutional neural networks (CNNs). More recently, models such as SASRec \cite{kang2018self} and BERT4Rec \cite{sun2019bert4rec} are built upon the self-attention mechanism \cite{vaswani2017attention, chen2024sigformer}, which automatically assigns weights to each interaction to capture their relative importance. Additionally, DROS \cite{yang2023generic} incorporates distributionally robust optimization (DRO) \cite{rahimian2019distributionally, wu2023understanding} to improve the model's robustness to out-of-distribution scenarios, which are common in RS \cite{wang2024distributionally, lin2024recommendation, chen2021autodebias, zhao2022popularity, gao2023alleviating, gao2023cirs}. The readers may refer to the survey \cite{fang2020deep,wang2019sequential} for more details.

% Please add the following required packages to your document preamble:
% \usepackage{booktabs}
% \usepackage{graphicx}
\begin{table}[t]
\centering
\caption{Time of constructing trie tree.}
\label{tab:time_trie}
{%
\begin{tabular}{@{}lcccc@{}}
\toprule
Dataset                          & Toy   & Book  & Clothing & Office \\ \midrule
Time & 0.38s & 0.17s & 0.81s    & 0.23s  \\ \bottomrule
\end{tabular}%
}
\end{table}

\subsection{LLMs for Recommendation}
Large language models (LLMs), with their powerful comprehension capabilities and extensive knowledge \cite{dubey2024llama, achiam2023gpt, wang2024llm4dsr}, have been widely applied in RS. 
% Their applications span various tasks, including collaborative filtering \cite{zhu2024collaborative, wang2024enhanced}, sequential recommendation \cite{hou2023learning, hou2022towards, bao2023bi}, graph-based recommendation \cite{wei2024llmrec}, and click-through rate (CTR) tasks \cite{bao2023tallrec, wang2023alt, xu2024enhancing}. Furthermore, LLMs have been utilized for tasks that were challenging for traditional models, such as denoising, explaining item embeddings \cite{tennenholtz2023demystifying}, interpreting recommendation results \cite{gao2023chat, wang2024can}, and enabling conversational recommendations \cite{gao2023chat}.
Two primary paradigms for using LLMs in RS as following.

\textbf{LLM-based Recommenders.} This paradigm attempts directly leveraging pre-trained LLMs as the backbone for recommendations using their zero-shot capabilities \cite{gao2023chat, hou2024large,liu2024once,wang2023zero,wang2024recommend}. However, these methods often perform poorly due to the significant discrepancy between the recommendation tasks and the training objectives of LLMs \cite{yang2024psl}. To address this, subsequent research has reformulated recommendation data into prompt formats and fine-tuned LLMs to improve their performance, achieving better results \cite{bao2023bi,bao2023tallrec,hu2024exact,li2023prompt, li2024citation,li2023exploring,li2024pap,shi2024enhancing,wang2024towards,zhang2023recommendation,zhang2024tired,gao2024end, lin2024bridging, xu2024slmrec, wang2023alt, wang2024enhanced, zhu2024collaborative}. 

However, these studies all rely on the LML for fine-tuning, without addressing the inherent misalignment between LML and recommendation tasks. To address this issue, S-DPO \cite{chen2024softmax} builds upon DPO \cite{rafailov2024direct} by constructing positive and negative samples, explicitly incorporating ranking information into the model training process. However, it suffers from performance instability, limited effectiveness, and low efficiency. Our proposed MSL effectively addresses these issues and significantly improves the performance of LLM-based RS.

\textbf{LLM-enhanced Recommenders.} This paradigm primarily utilizes LLMs in auxiliary roles, such as encoders to embed the semantic information of users and items \cite{xi2023towards, ren2024representation, wei2024llmrec, wang2023alt, lee2024star, ren2024enhancing, xu2024enhancing}, as an additional knowledge base \cite{yang2024common, qin2024d2k}, as a reasoning tool to generate chain-of-thought (CoT) data \cite{sun2024large, wang2024can}, or serves as teachers using distillation \cite{cui2024distillation, liu2024llm}. The main challenge of this paradigm lies in the significant gap between LLMs and traditional recommendation models, which hinders the effective transfer of knowledge.

\section{Conclusion}
% In this paper, we propose a novel loss function, MSL, specifically tailored for LLM-based RS. MSL leverages a Trie tree to identify valid tokens for each token-wise sample, effectively excluding invalid tokens from the loss calculation. While MSL offers significant advantages, it can lead to gradient vanishing issues during training. To mitigate this, we introduce an Adaptive Temperature Strategy, which dynamically adjusts the temperature coefficient to stabilize training. Compared to the original loss function, LML, designed for general LLMs, MSL demonstrates a closer alignment with the objectives of recommendation tasks. Experimental results across multiple datasets show that MSL significantly improves the performance of LLM-based recommenders while maintaining high computational efficiency.
In this paper, we introduce a novel loss function, MSL, specifically tailored for LLM-based RS. MSL excludes invalid tokens from participating in the loss calculation, achieving better alignment with the recommendation objectives and avoid the interference from
erroneous negative signal. Despite its advantages, it can lead to gradient vanishing issues during training. To mitigate this, we introduce the temperature coefficient and propose an Adaptive Temperature Strategy, which adaptively adjusts the temperature without requiring extensive hyperparameter tuning. We validate the effectiveness of MSL through theoretical analysis and empirical experiments. Our findings demonstrate that MSL significantly improves the performance of state-of-the-art LLM-based recommendation models.

This study highlights the importance of optimizing LLMs for recommendation tasks by refining their loss functions. Future research could explore the design of specialized LLM architectures to further enhance their suitability for recommendation systems.
%%
%% The acknowledgments section is defined using the "acks" environment
%% (and NOT an unnumbered section). This ensures the proper
%% identification of the section in the article metadata, and the
%% consistent spelling of the heading.
% \begin{acks}
% To Robert, for the bagels and explaining CMYK and color spaces.
% \end{acks}

%%
%% The next two lines define the bibliography style to be used, and
%% the bibliography file.

\begin{acks}
 This work is supported by the National Natural Science Foundation of China (62372399, 62476244), OPPO Research Fund, and the advanced computing resources provided by the Supercomputing Center of Hangzhou City University.
\end{acks}

%%
%% If your work has an appendix, this is the place to put it.
\appendix
\section*{Appendix}

\section{The proof of lemma \ref{lemma_ndcg}}
\label{apd:lemma_ndcg}

\begin{proof}
\begin{equation}
\begin{aligned}
\mathcal{L}_{MSL}(x,y^p;\theta) 
&= \sum_{t=1}^{|y^p|} - \log \frac{\exp(f_\theta(y_t^p))}{\sum_{z \in \mathcal{Z}_{valid}} \exp(f_\theta(z))} \\
&= - \log \prod_{t=1}^{|y^p|} P_{\theta}^{valid} \left(y^p_t \mid x, y^p_{<t}\right) = - \log P_{\theta}^{valid} \left(y^p \mid x \right) \\
\end{aligned}
\end{equation}
% \begin{equation}
% \begin{aligned}
% \mathcal{L}_{mask}(x,y^p;\theta) = \sum_{t=1}^{|y^p|} - \log \frac{\exp(f_\theta(y_t^p)))}{\sum_{z \in \mathcal{Z}_{valid}} \exp(f_\theta(z)))} = - \log P_{\theta}^{valid} \left(y^p \mid x \right)
% \end{aligned}
% \end{equation}

Since the probabilities of all items sum to 1,
% \begin{equation}
% \begin{aligned}
% - \log P_{\theta}^{valid} \left(y^p \mid x \right)
% &= - \log \frac{e^{\log P_{\theta}^{valid} \left(y^p \mid x \right)}}{\sum_{v \in \mathcal{V}} e^{\log P_{\theta}^{valid} \left(y^v \mid x \right)}} \\
% &= - \log \frac{1}{\sum_{v \in \mathcal{V}} e^{\log P_{\theta}^{valid} \left(y^v \mid x \right) - \log P_{\theta}^{valid} \left(y^p \mid x \right)}} \\
% \end{aligned}
% \end{equation}
\begin{equation}
\begin{aligned}
- \log P_{\theta}^{valid} \left(y^p \mid x \right) = - \log \frac{1}{\sum_{v \in \mathcal{V}} e^{\log P_{\theta}^{valid} \left(y^v \mid x \right) - \log P_{\theta}^{valid} \left(y^p \mid x \right)}}
\end{aligned}
\end{equation}

Using the inequality \(\exp(a) \geq \mathbb{I}(a), a \in \mathbb{R}\),

\begin{equation}
\begin{aligned}
\mathcal{L}_{MSL}(x,y^p;\theta) 
&\geq - \log \frac{1}{\sum_{v \in \mathcal{V}} \mathbb{I}(\log P_{\theta}^{valid} \left(y^v \mid x \right) \geq \log P_{\theta}^{valid} \left(y^p \mid x \right))} \\
&= - \log \frac{1}{\pi(p)}
\end{aligned}
\end{equation}
where $\pi(p)$  denotes the rank position of positive item $p$. Since \(\log(1 + \pi(p)) \leq \pi(p)\), 
\begin{equation}
\begin{aligned}
 - \log \frac{1}{\pi(p)} \geq - \log \frac{1}{\log(1+\pi(p))} = - \log NDCG(S)
\end{aligned}
\end{equation}
Since the denominator of the softmax function in each term of \(\mathcal{L}_{{MSL}}\) is smaller than \(\mathcal{L}_{LML}\), it follows that \(\mathcal{L}_{MSL} \leq \mathcal{L}_{LML}\), then
\begin{equation}
\begin{aligned}
 -\log NDCG(S) \leq \mathcal{L}_{MSL}  \leq \mathcal{L}_{LML}
\end{aligned}
\end{equation}

\end{proof}

\section{The proof of lemma \ref{lemma_ada_tau}}
\label{apd:lemma_adatau}
\begin{proof}
The denominator of \(P^{valid}_{\theta}(y_t^p)\), \(\mathbb{E}_f[\exp(f / \tau_t)]\), can be expanded as follows:
\begin{equation}
\begin{aligned}
&\mathbb{E}_f[\exp(f / \tau_t)]= \int_{-\infty}^{+\infty} \exp(f/\tau_t) \frac{1}{\sqrt{2\pi}\sigma_t} \exp\left(-\frac{(f-\mu_t)^2}{2\sigma_t^2}\right) df \\
% =& \int_{-\infty}^{+\infty} \frac{1}{\sqrt{2\pi}\sigma_t} \exp\left(-\frac{1}{2\sigma_t^2} ((f-\mu_t)^2-\frac{2\sigma_t^2f}{\tau_t})\right) df \\
=& \int_{-\infty}^{+\infty} \frac{1}{\sqrt{2\pi}\sigma_t} \exp\left(-\frac{1}{2\sigma_t^2} (f-(\mu_t+\frac{\sigma_t^2}{\tau_t}))^2 + \frac{\sigma_t^2}{2\tau_t^2} + \frac{\mu_t}{\tau_t}\right) df \\
=& \int_{-\infty}^{+\infty} \frac{1}{\sqrt{2\pi}\sigma_t} \exp\left(-\frac{(f-(\mu_t+\frac{\sigma_t^2}{\tau_t}))^2}{2\sigma_t^2}\right) df \exp(\frac{\mu_t}{\tau_t}) \exp(\frac{\sigma_t^2}{2\tau_t^2}).
\end{aligned}
\end{equation}

The integral corresponds to a standard Gaussian distribution is 1. Therefore,
\begin{equation}
\mathbb{E}_f[\exp(f / \tau_t)] = \exp(\mu_t / \tau_t) \exp(\sigma_t^2/2\tau_t^2).
\end{equation}

Thus, we approximate:
\begin{equation}
\mathbb{E}_f[\exp(f / \tau_t)] \approx \exp(\mu_t / \tau_t) \left(1 + \frac{\sigma_t^2}{2 \tau_t^2}\right),
\end{equation}
since \(\exp(\sigma_t^2 / 2 \tau_t^2) = 1 + \frac{\sigma_t^2}{2 \tau_t^2} + o((\sigma_t^2 / 2 \tau_t^2)^2)\). Then the original equation can be approximated as:
\begin{equation}
P^{valid}_{\theta}(y_t^p) \approx \frac{\exp(f_\theta(y_t^p) / \tau_t)}{|\mathcal{Z}_{valid}| \exp(\mu_t / \tau_t + \sigma_t^2 / 2 \tau_t^2)} = \eta.
\end{equation}
Solving the above quadratic equation, 
\begin{equation}
\tau_t \approx \frac{(f_\theta(y_t^p) - \mu_t) - \sqrt{(f_\theta(y_t^p) - \mu_t)^2 - 2 \sigma_t^2 \log(|\mathcal{Z}_{valid}| \eta)}}{2 \log(|\mathcal{Z}_{valid}| \eta)}
\end{equation}
\end{proof}

% The denominator of \(P^{valid}_{\theta}(y_t^p)\), \(\mathbb{E}_f[\exp(f / \tau_t)]\), can be expanded using a Taylor series as follows:
% \begin{align}
% \mathbb{E}_f[\exp(f / \tau_t)] &= \exp(\mu / \tau_t) \\
% &\left( 1 + \frac{\mathbb{E}_f\left[ (f-\mu)\right]}{\tau_t} 
% + \frac{\mathbb{E}_f\left[ (f-\mu)^2\right]}{2\tau_t^2} 
% + \sum_{i=3}^{\infty}\frac{\mathbb{E}_f\left[ (f-\mu)^i\right]}{i!\tau_t^i} \right)
% \end{align}
% Since the higher-order terms are bounded, we have:
% \begin{equation}
% \left|\sum_{i=3}^{\infty}\frac{\mathbb{E}_f\left[ (f-\mu)^i\right]}{i!\tau_t^i} \right| \leq \sum_{i=3}^{\infty} \frac{2(\lambda/2)^i i!}{\tau_t^i i!} = \frac{2(\frac{\lambda}{2\tau_t})^3}{1-\frac{\lambda}{2\tau_t}} \leq \frac{1}{24}
% \end{equation}
% where we use the fact that the central moment of a sub-exponential variable is bounded:
% \begin{align}
% \mathbb{E}_f\left[|f-\mu|^i\right] &= \int_0^{\infty} p\left(|f-\mu|^i > t\right) dt = \int_0^{\infty} p\left(|f-\mu| > t^{1/i}\right) dt \\
% &\leq \int_0^{\infty} 2 e^{-\frac{2 t^{1/i}}{\lambda}} dt 
% = 2(\lambda / 2)^i i \int_0^{\infty} e^{-u} u^{i-1} du 
% = 2(\lambda / 2)^i i!
% \end{align}
% Thus, we approximate:
% \begin{equation}
% \mathbb{E}_f[\exp(f / \tau_t)] \approx \exp(\mu / \tau_t) \left(1 + \frac{\sigma^2}{2 \tau_t^2}\right)
% \end{equation}
% Since \(\exp(\sigma^2 / 2 \tau_t^2) = 1 + \frac{\sigma^2}{2 \tau_t^2} + o((\sigma^2 / 2 \tau_t^2)^2)\) and \(\frac{\sigma^2}{2 \tau_t^2} \leq \frac{\lambda^2}{2 \tau_t^2} \leq \frac{1}{8}\), the original equation can be rewritten as:

\section{Validation of Gaussian distribution for logits}
\label{apd:gaussian}
\begin{figure}[h]
  \centering
  \includegraphics[width=\linewidth]{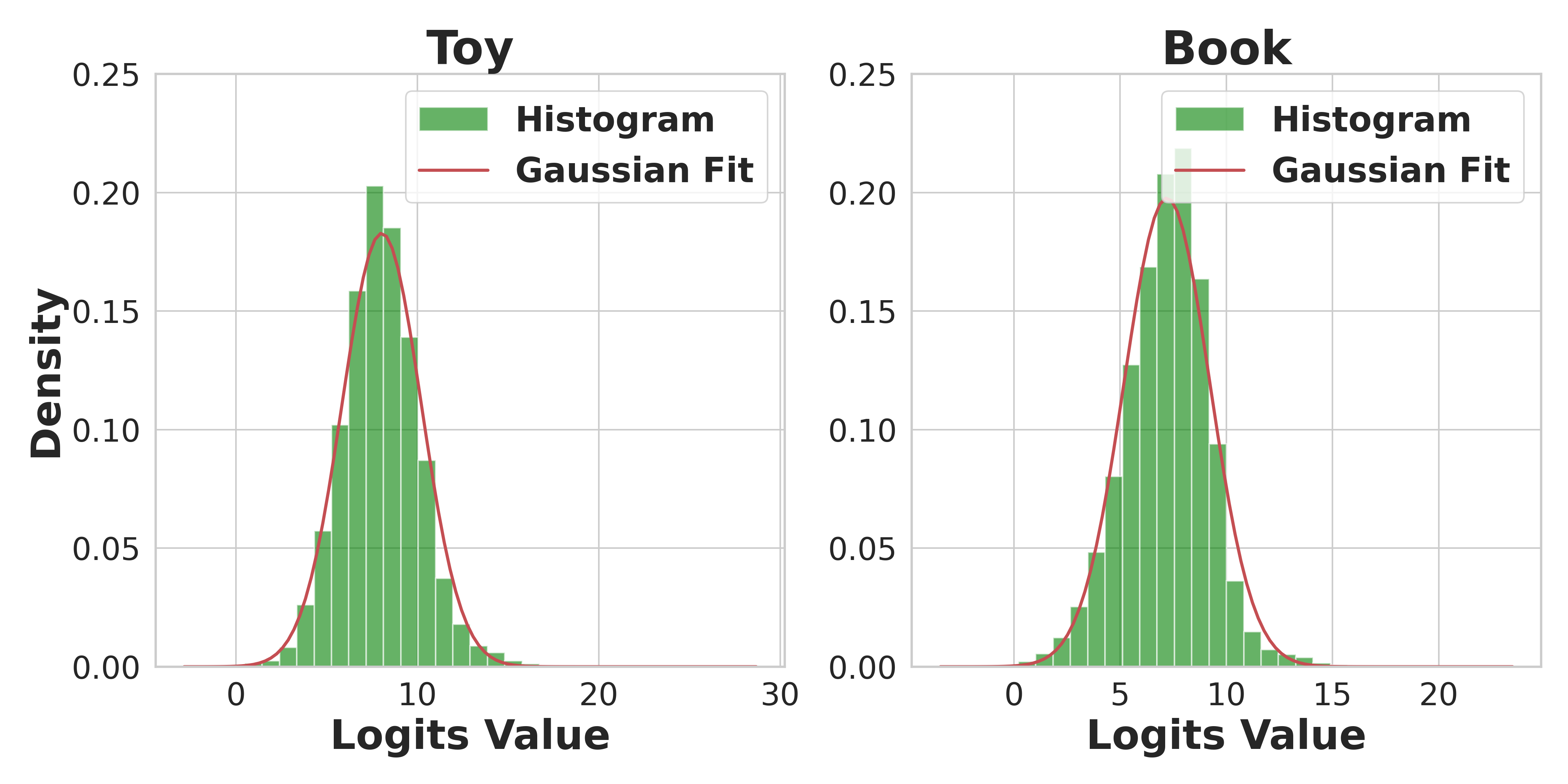}
  \caption{The logits distribution and the fitting results using a Gaussian distribution on Toy and Book dataset.}
  \Description{}
  \label{logits_gaussian_fit}
\end{figure}

% Figure \ref{logits_gaussian_fit} shows the logits distribution for a batch of samples from two datasets, along with their Gaussian fitting curves. The results indicate that the logits distribution closely follows a Gaussian distribution.
Figure \ref{logits_gaussian_fit} indicates that the logits distribution closely follows a Gaussian distribution.

\bibliographystyle{ACM-Reference-Format}
\bibliography{sample-base}

\end{document}